\documentclass[graybox, envcountchap]{svmult}

\newcommand{\apj}{Astrophys.\ J.\ }
\newcommand{\apjl}{Astrophys.\ J.\ }
\newcommand{\apjs}{Astrophys.\ J. Suppl.\ }
\newcommand{\aap}{Astron.\ Astrophys.\ }

\newcommand{\prd}{Phys.\ Rev.\ D\ }

\newcommand{\araa}{Annu.\ Rev.\ Astron. Astrophys.\ }

\newcommand{\nat}{Nature~}

\newcommand{\mnras}{Mon.\ Not.\ R.\ Astron.\  Soc.\ }
\newcommand{\pasj}{Publ.\ Astron.\ Soc.\ Japan\ }
\newcommand{\apss}{Astrophys.\ Space Sci.\ }

\newcommand{\jqsrt}{J.\ Quant.\ Spectrosc.\ Radiat.\ Transfer\ }

\newcommand{\nar}{New Astron.\ Rev.\ }

\usepackage{mathptmx}        
\usepackage{amsmath}
\usepackage{amssymb}
\usepackage{color}
\usepackage[T1]{fontenc}
\usepackage[utf8]{inputenc}
\usepackage{helvet}          
\usepackage{courier}         
\usepackage{dirtree}

\usepackage{makeidx}        
\usepackage{graphicx}        
\usepackage{subfig}

\usepackage{multicol}        
\usepackage[bottom]{footmisc}

\usepackage{hyperref}        
\hypersetup{colorlinks=true,urlcolor=blue}

\usepackage[misc]{ifsym}

\usepackage[numbers]{natbib}

\makeindex             

\begin{document}


\title{Polarization Signatures from GRMHD Simulations of Black Hole Accretion}
\titlerunning{Polarization Signatures from GRMHD Simulations of Black Hole Accretion}
\author{P. Chris Fragile, Maciek Wielgus and Cora Prather}
\institute{P. Chris Fragile (\Letter) \at College of Charleston, 66 George Street, Charleston, SC 29424 USA, \email{fragilep@cofc.edu} \\
Center for Computational Astrophysics, Flatiron Institute, 162 5th Avenue, New York, NY 10010 USA
\and Maciek Wielgus \at Instituto de Astrofísica de Andalucía-CSIC, Glorieta de la Astronomía s/n, E-18008 Granada, Spain, \email{maciek@wielgus.info} \\
Research Centre for Computational Physics and Data Processing, Institute of Physics, Silesian University in Opava, Bezru\v{c}ovo n\'am.~13, CZ-746\,01 Opava, Czech Republic
\and Cora Prather \at Black Hole Initiative at Harvard University, 20 Garden Street, Cambridge, MA 02138, USA, \email{cprather@fas.harvard.edu}}
%
%
\maketitle

\abstract{This chapter tells the still-unfolding story of extracting polarization signatures from general relativistic magnetohydrodynamics simulations of accretion disks. In some sense, this effort is premature as there are still very few results of this kind. Much more abundant are phenomenological models. Nevertheless, we feel now is the time to rally the community to this cause. Since the focus of this book is on X-ray polarimetry, we focus exclusively on simulations of accretion onto compact objects. Most of the relevant work so far has been on black hole accretion disks, though neutron stars are also viable targets for X-ray polarimetry. The focus of our chapter is on how X-ray polarimetry coupled with accretion simulations might help us better understand properties of the disks, coronae, and jets that are the dominant components of accreting compact sources. We briefly illustrate the promise of this technique by demonstrating how it has already been used in the case of the Event Horizon Telescope (using radio polarimetry). We also speculate about where this field may be heading in the near future.}


\section{Background}
\label{sec:background}

We now have a decades long history of hydrodynamic and magnetohydrodynamic (MHD) simulations of black hole accretion disks. There are simulations of thin disks \cite{Shafee08, Penna10, Penna12, Sadowski16, Lancova2019, Mishra22}, thick disks \cite{DeVilliers03, Gammie03, Narayan12, Shiokawa12, Yuan12, Sadowski13a}, disks at low \cite{Dibi12, Sadowski17, Jiang19, Dexter21} and high \cite{Jiang14, Sadowski14, McKinney14, Utsumi22, Yoshioka22, Fragile25, Zhang25} mass accretion rates, disks with weak \cite{Beckwith08} and strong \cite{Tchekhovskoy11, McKinney12, Wielgus2015, White19a, Curd23, Scepi24} magnetic fields, and tilted \cite{Fragile07, McKinney13, Morales14, Liska18, Liska19, White19b} and truncated \cite{Takahashi16, Liska22a, Bollimpalli24} disks. There are simulations that include non-ideal MHD effects \cite{Dionys2013, Foucart2016} and non-Kerr black holes \citep{Mizuno2018,Chatterjee2023}. For a partial review, see \cite{Abramowicz13}. The point is, there are many existing simulations of black hole accretion disks with more being published every day.

Some of these simulations have included radiative processes during the evolution \cite{Jiang14, Sadowski14, McKinney2017, Sadowski17, Dexter21, Utsumi22, Yoshioka22, Fragile25, Zhang25}, while others have coupled their output to radiative post-processing codes \cite{Davis09, Dolence09, Schnittman13b, Dexter16, Narayan16, Bronzwaer18, Roth25}. Through these means, researchers have tried to extract products from the simulations akin to what observers might measure from real sources. These efforts have variously focused on: creating spectra \cite{Schnittman13a, Kinch19, Wielgus2022, Mills24, Moscibrodzka2024, Roth25}, images \cite{Shcherbakov12, Chan15, EHTC_M87_p5, EHTC_SgrA_p5}, and line profiles \cite{Kinch16}; measuring the radiative efficiency \cite{Noble09, Kinch21}; and characterizing time variability \cite{Schnittman06a, Schnittman06b, Mishra17, Bollimpalli20,Salas2025}.

Another window into black hole accretion is polarization. For many decades now this has been done at optical \cite{Baade1956}, infrared \cite{Angel1980}, and radio \cite{Wardle1974} wavelengths. The point of this book is that polarization studies of black hole accretion are now also possible in X-rays \cite{Doroshenko22, Krawczynski22, Podgorny23}, through the Imaging X-ray Polarimetry Explorer \citep[IXPE;][]{Weisskopf22}, the X-ray Polarimeter Satellite \citep[XPoSat;][]{Radhakrishna25}, and hopefully continuing with proposed future missions such as the enhanced X-ray Timing and Polarimetry mission \citep[eXTP;][]{Zhang19}. Among other potential discoveries, these missions can probe the geometries of accretion disks \cite{Schnittman09, Cheng16, Farinelli23} and coronae \cite{Schnittman10, Marinucci22, Gianolli23}, as well as constrain the black hole spin parameter \citep{Marra2024}. This has motivated new work to try to connect numerical GRMHD to real systems through comparison of polarization properties, an approach that is more physically self-consistent than using semi-analytic models \cite{Schnittman09,Saurabh2025}.

In the next section we describe in general terms the techniques involved in extracting polarization information from black hole accretion simulations (Section \ref{sec:techniques}). We break the discussion into two parts, first reviewing some details of the simulations themselves (Section \ref{sec:simulations}), then presenting a discussion about the process of turning the output of simulations into polarization products (Section \ref{sec:transport}). That section also discusses some of the polarization-capable radiation transport codes in use today. After our review of the techniques, we discuss some of the results that have been published so far, dividing our discussion roughly by where the polarized radiation originates: the disk (Section \ref{sec:disk}), the corona (Section \ref{sec:corona}), and the jet (Section \ref{sec:jet}). We also discuss the role that polarization may be able to play in understanding black hole spectral states (Section \ref{sec:states}) and the associated time variability (Section \ref{sec:variability}). Then, although this book is focused on X-ray polarimetry, we include a section on the connections between black hole accretion simulations and polarization in the context of the Event Horizon Telescope (Section \ref{sec:EHT}). Finally, in Section \ref{sec:conclusion}, we give our concluding thoughts and propose possible future directions for the community.

\section{Techniques \& Codes}
\label{sec:techniques}

The basic approach of extracting polarization signatures from black hole accretion disk simulations is pretty straightforward. The simulations evolve and ultimately store the relevant plasma properties (density, temperature, velocity, magnetic field, etc.) as a function of spatial location at discrete time intervals. This information is then passed to a polarization-capable radiation transport code to produce spectra, light curves, and polarization maps analogous to what would be observed by IXPE, XPoSat, or other polarization missions. In the rest of this section, we delve into more of the details about how each of these steps -- the simulations (\ref{sec:simulations}) and the radiation transport (\ref{sec:transport}) -- are done.

\subsection{Simulation Techniques \& Codes}
\label{sec:simulations}

Most simulations of black hole accretion disks these days solve the following set of conservation equations for general relativistic hydrodynamics (GRHD):
\begin{eqnarray}
 \partial_\lambda \left(\sqrt{-g}\rho u^\lambda\right) &=& 0 ~,  \label{eq:de} \\
 \partial_\lambda \left(\sqrt{-g}~T^\lambda_\mu\right) &=& \sqrt{-g}~T^\alpha_\beta~\Gamma^\beta_{\mu \alpha} ~,
    \label{eq:en}
\end{eqnarray}
where $g$ is the metric determinant, $\rho$ is the fluid rest-mass density, $u^\mu$ is the fluid 4-velocity, and
\begin{equation}
T^{\alpha \beta}_\mathrm{FLU} = (\rho + \rho \epsilon + P_\mathrm{gas}) u^\alpha u^\beta + P_\mathrm{gas} g^{\alpha \beta} ~,
\end{equation}
is the fluid stress-energy tensor for a gas pressure $P_\mathrm{gas}$ and specific internal energy $\epsilon$. For an adiabatic equation of state, $P_\mathrm{gas} = (\Gamma-1)\rho\epsilon$,
where $\Gamma$ (without subscripts or superscripts) is the adiabatic index. With indices, $\Gamma^\alpha_{\beta\gamma}$ indicates the geometric connection coefficients of the metric (a.k.a. the Christoffel symbols).

Most simulations also simultaneously evolve a magnetic field (so magnetohydrodynamics or GRMHD). In the ideal approximation (where the electric fields vanish in the fluid frame), the homogeneous Maxwell equations can be written 
\begin{equation}
 \partial_\lambda \left(\sqrt{-g} F^{\lambda j}\right) = 0 ~, 
      \label{eq:ind} 
\end{equation}
where $F^{\alpha\beta}$ is the Faraday tensor. Including the magnetic field also requires modifying the stress-energy tensor to
\begin{equation}
T^{\alpha \beta}_\mathrm{MHD} = (\rho + \rho \epsilon + P_\mathrm{gas} + 2P_\mathrm{mag}) u^\alpha u^\beta + (P_\mathrm{gas} + P_\mathrm{mag}) g^{\alpha \beta} - b^\alpha b^\beta ~,
\end{equation}
where $P_\mathrm{mag} = b_\alpha b^\alpha/2$ is the magnetic pressure, $b^\alpha$ is the magnetic field 4-vector, which is defined as the dual of the Faraday tensor projected onto the fluid 4-velocity $b^\alpha \equiv u_\beta {^*F^{\alpha\beta}}$.
Along with the evolution equation (\ref{eq:ind}), the magnetic field 3-vector $B^j \equiv {^*F^{tj}}$ should also satisfy Gauss's law of magnetism (or the solenoidal constraint) $\partial_j (\sqrt{-g}B^j) = 0$. Most GRMHD codes these days take some care to start with a divergence-free magnetic field and evolve it in a way that it remains so \citep[see][for further discussion]{Toth00}. 

More recently, some black hole accretion simulations have been run with radiation included self-consistently in the evolution, or radiation MHD (GRRMHD). This is especially important whenever the luminosity of the accretion disk approaches or exceeds the Eddington limit $L_\mathrm{Edd} \approx 10^{38} M/M_\odot$ erg s$^{-1}$. Here there are two general approaches. One is to use a moment formalism \citep{Thorne81, Shibata11} to write the radiation transport as a set of conservation equations \cite{Anninos20}
\begin{eqnarray}
 \partial_\lambda \left[\sqrt{-g} R^\lambda_{\mu (\nu)}\right] & = &
     \frac{\partial}{\partial\nu}\left[\nu M_{\mu\beta\gamma (\nu)} u^{\beta ;\gamma}\right] \\
     & &+ \sqrt{-g}~R^\alpha_{\beta (\nu)}~\Gamma^\beta_{\mu \alpha} - \sqrt{-g}~G_{\mu (\nu)} 
    \label{eq:rad_en} 
\end{eqnarray}
for the radiation stress-energy
\begin{equation}
R^{\alpha \beta}_{(\nu)} = J_{(\nu)} u^\alpha u^\beta + H^\alpha_{(\nu)} u^\beta + H^\beta_{(\nu)} u^\alpha + L^{\alpha \beta}_{(\nu)} ~,
\end{equation}
where $ G_{\alpha (\nu)}$ represents radiation-matter interaction source terms, $M^{\alpha\beta\gamma}_{(\nu)}$
is the third-rank moment tensor associated with Doppler and gravitational frequency shifts, and $J_{(\nu)}$, $H^\alpha_{(\nu)}$, and $L^{\alpha \beta}_{(\nu)}$ are the radiation energy density, momentum density, and stress tensor, respectively, in the fluid frame. The $(\nu)$ subscripts indicate frequency-dependent quantities. Moment methods produce equations that are very similar to the fluid conservations equations (e.g., \ref{eq:en}), but likewise require a closure relation, with M$_1$ \cite{Levermore84} being the most popular for GRRMHD \cite{Sadowski13b, Fragile14, McKinney14}. 

The other option is to solve the full Boltzmann transport equation, which can be written in various forms \cite{Davis20}, including
\begin{eqnarray}
\frac{\partial}{\partial t} \left[n^t n_\mu I_{(\nu)}\right] + \frac{1}{\sqrt{-g}} \frac{\partial}{\partial x^i} \left[\sqrt{-g} n^i n_\mu I_{(\nu)}\right] + \frac{\partial}{\partial \nu} \left[n_\nu n_\mu I_{(\nu)}\right] \\
-\frac{1}{\sin \zeta} \frac{\partial}{\partial \zeta}\left[n_\zeta n_\mu I_{(\nu)}\right] + \frac{\partial}{\partial \psi}\left[n_\psi n_\mu I_{(\nu)}\right] + n^\alpha n_\delta \Gamma^\delta_{\mu\alpha} I_{(\nu)} \\
= n_\mu \left[j_{(\nu)} - \alpha_{(\nu)} I_{(\nu)}\right] ~,
\end{eqnarray}
where $I_{(\nu)}$ is the radiation intensity, $\zeta$ and $\psi$ are the polar and azimuthal solid angles, respectively, $j_{(\nu)}$ is the emissivity, and $\alpha_{(\nu)}$ is the extinction coefficient. 

Codes using moment methods can generally perform multi-dimensional simulations of black hole accretion faster than codes that perform the full radiation transport. However, with ever larger and faster computing platforms becoming available, the extra cost of solving full Boltzmann transport is becoming more acceptable \cite{Zhang25}.  

Although we retained the frequency dependence in the radiation equations above, most GRRMHD simulations to date have been done using the ``gray'' approximation, where the frequency dependence is integrated out and only a frequency-averaged representations of the radiation are evolved. This is again done to reduce the computational cost, although some recent work has begun to relax this approximation and explore multi-frequency radiation transport, where spectral information is retained, at least in some limited number of frequency bins \cite{Fragile23, Mills24}.

There are now a number of codes available or in development to solve the equations of GRMHD or GRRMHD \citep{Anninos05, DelZanna07, Sadowski13b, Takahashi16, White16, Porth17, Prather21, Liska22b}. A comparison of the major codes used for black hole accretion simulations is presented in \citep{Porth19}. All of these techniques and codes evolve some variant of the conserved fields $\mathbf{U} = \{\sqrt{-g}(\rho u^t, ~T^t_t, ~T^t_j, ~F^{ti}, ...)\}$ according to evolution equations like the ones described above. After each update step, though, these conserved fields must be converted into a corresponding set of primitive fields  $\mathbf{P}=\{\rho, ~\rho\epsilon, ~u^i, ~B^i,...\}$. These are usually the fields that need to be fed into the radiation transport codes, as described in the next subsection. How frequently these primitive fields are saved to file plays a role in determining whether the radiation transport code must treat the disk as a static structure (if the dumps are too infrequent) or can account for the dynamics by ingesting many sequential time slices.

\subsection{Radiation Transport Techniques \& Codes}
\label{sec:transport}

In the absence of scattering and refraction, light in the curved spacetime around a compact object follows null geodesics, which can be computed for a given spacetime through a set of four coupled ordinary differential equations parameterizing the four-position $x^\mu$ along the geodesic path:
\begin{equation}
    \label{eqn:geo}
    \frac{d^2 x^\mu}{ds^2} = - \Gamma^\mu_{\nu \lambda} \frac{d x^\nu}{ds} \frac{d x^\lambda}{ds} ~,
\end{equation}
where $s$ is an affine parameter.  These equations can be followed forward to compute the propagation of emitted light or backward to determine the lines of sight from some perspective or camera. Analytic solutions to these equations exist for the Schwarzschild and Kerr spacetimes \cite{Dexter2009, Gralla2020}.  However, due to the complexity of these solutions, or for generalization to other metrics, some codes opt instead to solve the ordinary differential equations numerically, e.g., with Runge-Kutta or symplectic methods.

In the Newtonian case (or equivalently in the rest frame of the fluid) the polarized radiation transport equation reads:
\begin{equation}
\label{eqn:pol_rad}
\frac{d}{d s}
\begin{pmatrix} I \\ Q \\ U \\ V \end{pmatrix}
 = \begin{pmatrix} j_{I} \\ j_{Q} \\ j_{U} \\ j_{V} \end{pmatrix}
- \begin{pmatrix} 
\alpha_{I} & \alpha_{Q} & \alpha_{U} & \alpha_{V} \\
\alpha_{Q} & \alpha_{I} & \rho_{V} & -\rho_{U} \\
\alpha_{U} & -\rho_{V} & \alpha_{I} & \rho_{Q} \\
\alpha_{V} & \rho_{U} & -\rho_{Q} & \alpha_{I} 
\end{pmatrix}
\begin{pmatrix} I \\ Q \\ U \\ V \end{pmatrix} ~,
\end{equation}
where the coefficients $\{j_{I}, j_{Q}, j_{U}, j_{V}\}$ are the specific emissivities at the desired frequency corresponding to each Stokes parameter, $\{\alpha_{I}, \alpha_{Q}, \alpha_{U}, \alpha_{V}\}$ are the absorption coefficients, and $\{\rho_{Q}, \rho_{U}, \rho_{V}\}$ are the rotation coefficients, which map types and directions of polarization while conserving intensity. In order to accommodate general relativistic transport, two additional effects must be computed: the frame transformation from the fluid to the observer frame (which generates Doppler effects and gravitational redshift) and parallel transport of the direction of linear polarization through the intervening spacetime. The frame transformation can be accounted for as a frequency shift between the emitter and observer frames:
\begin{equation}
\nu\mathrm{(obs)} = \nu\mathrm{(em)} \frac{u^\nu\mathrm{(obs)}k_\nu\mathrm{(obs)}}{u^\mu\mathrm{(em)}k_\mu\mathrm{(em)}} ~,
\end{equation}
where $u^\mu$(em) and $k_\mu$(em) are the fluid four-velocity and photon four-momentum, respectively, in the emitted frame and $u^\nu$(obs) and $k_\nu$(obs) are the same in the observer frame. Whenever a photon scatters, the old observer frame becomes the new emitter frame. Between scattering (or other interaction) events, the polarization vector $p^\nu$ is parallel transported according to 
\begin{equation}
\frac{dp^\nu}{ds} = -\Gamma^\nu_{\mu\sigma} p^\sigma k^\mu/k^0 ~.
\end{equation}
We call attention to the fact that many other mathematically equivalent formulations for relativistic polarized transport exist \cite{Broderick2004, Gammie2012}.

Of particular interest to us are the Stokes parameters associated with linear polarization $\{I, Q, U\}$; this is because current-generation X-ray polarimeters do not measure circular polarization. It will be convenient for later discussion for us to introduce the polarization degree, $\delta$, which represents the fraction of the total intensity that is linearly polarized, and the polarization angle, $\psi$, which represents the orientation of the linear polarization vector. These can be recovered from the following relations to the Stokes parameters:
\begin{eqnarray}
Q/I & = \delta \cos 2\psi \nonumber \\
U/I & = \delta \sin 2\psi ~.
\end{eqnarray}
The polarization degree $\delta$ remains invariant along the geodesic, while the polarization angle $\psi$ can be rotated by the parallel transport.

Codes which simulate the propagation of light can be crudely separated into two types: 1) ray-tracing codes that are primarily interested in producing resolved images by following lines of sight for each element of a closely-spaced grid (pixels), resulting in a prediction for the emitted intensity distribution on the sky (an image), a method first used in \cite{Luminet79}; and 2) radiative modeling codes that capture the emission, absorption, and scattering of light along a geodesic and compute the resulting luminosity and spectral energy distribution. The former are mostly used for computing the appearance of resolvable compact objects, such as M87* \citep{EHTC_M87_p1} and Sgr A* \citep{EHTC_SgrA_p1}. Thus, they are primarily focused on radio frequencies to take advantage of Very-long Baseline Interferometry (VLBI) networks capable of actually resolving the necessary scales -- tens of micro-arcseconds ($\mu$as). Since this review is mostly about X-ray polarization, where spatial resolution of accretion disks is not possible, we focus exclusively on the latter type of code -- those aimed at producing spectra of intensity, polarization degree, and polarization angle.

For computing the full spectra of compact objects, Monte Carlo methods are not technically required but are universally adopted as being much more efficient than direct solution.  While direct solution is sometimes employed in transport during GRMHD simulations \cite{White2023}, the cost scales with the number of frequency bins, making it prohibitively expensive for general use.

Monte Carlo radiative modeling codes share nearly all of their properties and implementation with ray-tracing codes, with the exception that the starting points for ``superphotons'' representing emission are computed according to the local emissivity, instead of being chosen to land at a camera.  In addition, the inverse Compton scattering of photons quickly becomes important, and at high frequencies the bremsstrahlung emission from electrons as well \cite{Yarza2020}, so these are treated as scattering kernels and an enhanced emissivity, respectively.

A number of codes exist for predicting the spectra of unpolarized, fully ionized accretion flows \cite{Dolence09, Kawashima2023, Pelle2022}.  The code {\tt Pandurata} \cite{Schnittman13b,Kinch19} adds polarization and Fe K$\alpha$ line calculations, which {\tt MONK} \cite{Zhang19} emulates. {\tt RADPOL} \cite{Moscibrodzka2020} additionally supports non-Kerr spacetimes. 

\section{Disk}
\label{sec:disk}

We now begin to look at applications of the techniques described in the previous section. However, before considering the polarization signatures of full GRMHD simulations, we begin with a couple of simple analytic models. Our reasoning is that it will allow us to more cleanly separate the polarization signatures of different accretion components before considering more complicated simulation results. To begin, we consider an analytic model of a razor thin accretion disk \citep[mostly following][]{Schnittman09}. 

In Newtonian gravity, symmetry arguments demand that the polarization angle measured from such a disk must be either parallel or perpendicular to its symmetry axis. The only thing that can vary (with inclination) is the polarization degree $\delta$. By contrast, relativistic effects, such as aberration, beaming, gravitational lensing, and frame-dragging, can cause both the polarization degree and angle to vary across the disk in nontrivial ways \cite{Stark77, Connors80}. It has even been suggested that the resulting relativistic polarization signatures could be used to infer the spin of a black hole if it were observed in a disk-dominated state \cite{Laor90, Dovciak08}.

\subsection{Energy-integrated Images and Polarization Maps}

\begin{figure}
\sidecaption
\includegraphics[width=0.5\linewidth,trim=0mm 0mm 0mm 0,clip]{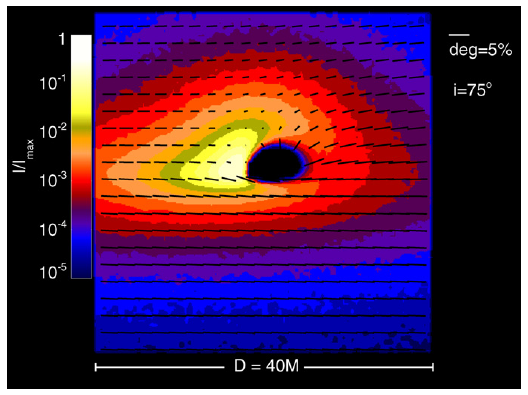}
\includegraphics[width=0.5\linewidth,trim=0mm 0mm 0mm 0,clip]{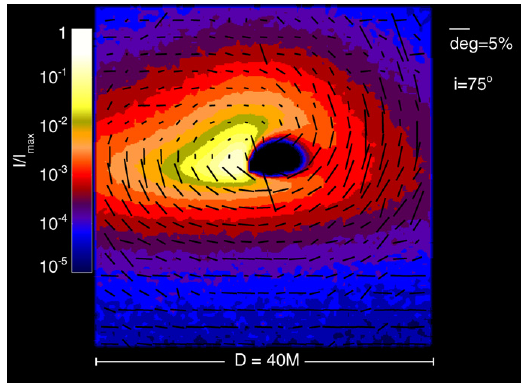}
%
%
\caption{{\em Left:} Ray-traced image of a thin Novikov-Thorne accretion disk with $L=0.1L_\mathrm{Edd}$ around a black hole with spin $a/M=0.99$. The observer is looking from an inclination of $75^\circ$ relative to the disk rotation and black hole spin axes, with the gas on the left side approaching the observer. The observed intensity is represented by the logarithmic color scale. The energy-integrated polarization at each location is represented by the length and orientation of the black ticks. {\em Right:} Same as the image on the left, except now including returning radiation (or reflection). Photons emitted from the inner disk have their trajectories bent by the black hole and scatter off the opposite side of the disk toward the distant observer. Image reproduced with permission from \cite{Schnittman09}, copyright by AAS.}
\label{fig:disk_image}       
\end{figure}

In Figure~\ref{fig:disk_image}, we show simulated images and polarization maps from \cite{Schnittman09} for a Novikov–Thorne accretion disk \cite{Novikov73} around a black hole with spin $a/M= 0.99$ and luminosity $L=0.1 L_\mathrm{Edd}$ viewed by an observer at an inclination of $i= 75^\circ$\footnote{We use the standard convention for inclination angle, with $i=0^\circ$ being an observation face-on to the disk and $i=90^\circ$ being edge-on.}. Before considering the polarization vectors, let us focus first on the disk images themselves. Numerous relativistic effects are clearly apparent. For one, there is increased intensity on the approaching side of the disk relative to the receding one due to special relativistic beaming. For another, we notice that general relativistic light bending (a.k.a. gravitational lensing) makes the far side of the disk appear warped and bent up toward the observer. These effects are well known from the first attempts to create synthetic images of black hole accretion disks \cite{Luminet79}.

On top of the intensity images in Figure~\ref{fig:disk_image} are superposed polarization maps with small black ticks representing the local energy-integrated polarization degree (via the length) and angle (via the orientation) of the radiation. We focus first on the left panel, which only includes photons traveling directly from the disk to the observer (ignoring so-called returning radiation). In parts of the disk far from the black hole, we see that the polarization ticks are oriented parallel to the disk plane ($\psi=0^\circ$) with lengths equal to about $\delta = 0.04$ (4\% polarization), as expected for a scattering-dominated atmosphere observed at this inclination \cite{Chandrasekhar60}. 

Nearer to the black hole, a number of relativistic effects alter the polarization. The two most prominent are the same ones we noted in the disk image: special relativistic beaming and general relativistic light bending. In the case of relativistic beaming, photons emitted nearly normal to the disk plane in the fluid frame are beamed into the direction of the local orbital motion as seen by a distant observer. When looking at the approaching side of the disk, the result is a smaller effective inclination and thus also a smaller degree of polarization, as is apparent on the left side of the disk in Figure~\ref{fig:disk_image} (left panel). On the other hand, photons emitted at large inclinations in the local fluid frame, but in a direction opposite the orbital motion will be beamed up toward observers at lower inclinations, resulting in a higher degree of polarization than what would otherwise be observed. This can be seen on the right side of the disk in Figure~\ref{fig:disk_image} (left panel). Because the intensity of the radiation is higher on the approaching side of the disk (where polarization is lowered), the net effect of relativistic beaming is to lower the overall polarization degree seen by an observer.

Gravitational lensing, as we have already seen, causes the far side of the disk to appear warped up toward the observer. This effectively lowers the inclination of the photons that reach the observer from that part of the disk, which lowers the polarization degree. This is apparent in the tick marks above the black hole in Figure~\ref{fig:disk_image} (left panel). Another impact of gravitational lensing is that photons can leave one side of the disk, and because of their curved trajectory near the black hole, strike the other side and then reflect to the observer. This so-called returning radiation \cite{Cunningham76} does not dramatically alter the disk image (compare left and right panels of Figure~\ref{fig:disk_image}), but it can have a substantial effect on the polarization, both in degree and angle. Some returning radiation starts from the far side of the disk (top of the image in Figure~\ref{fig:disk_image} right panel) and is reflected off the near side (bottom of the image) with a relatively small scattering angle. This radiation will retain its moderate horizontal polarization. On the other hand, photons emitted from one side (left or right) of the disk can be bent to the other side and then scatter at roughly $90^\circ$ to reach the observer. This large-angle scattering in a plane roughly parallel to the disk will produce a polarization that is nearly vertical ($\psi \approx \pm 90^\circ$) in Figure~\ref{fig:disk_image}. In other words, the polarization of the returning radiation is rotated at approximately a right angle with respect to the usual disk polarization angle. This can be seen on both the left and right sides of the disk in Figure~\ref{fig:disk_image} (right panel). Although relatively small in total flux, this contribution can dominate the total polarization, the reason being, a small flux with a very high polarization can contribute more to the total polarization than a high flux with a very small polarization.

Because photons from the inner parts of the disk experience stronger gravitational deflection, they are more likely to return to the disk than the photons emitted from larger radii. Thus, there is a characteristic ``transition radius,'' within which returning radiation dominates and produces a net vertical polarization. Outside this radius, the direct radiation dominates and produces horizontal polarization. Of course, where this transition radius is located depends sensitively on the spin of the black hole, which is one way polarization might be used to constrain this parameter \cite{Schnittman09,Marra2024}.

\subsection{Spectral Information}

\begin{figure}
\sidecaption
\includegraphics[width=1.0\linewidth,trim=0mm 0mm 0mm 0,clip]{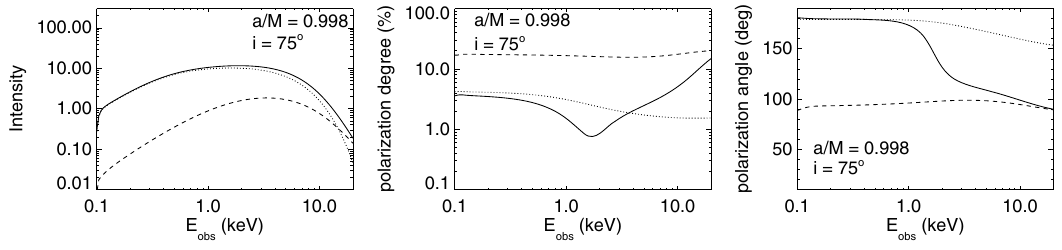}
%
%
\caption{{\em Left:} Observed intensity, {\em Middle:} degree of polarization, and {\em Right:} angle of polarization, including direct (dotted curves), returning (dashed curves), and total (solid curves) contributions, for a Novikov-Thorne disk with $L = 0.1 L_\mathrm{Edd}$ around a black hole with spin $a/M = 0.998$ for an observer at $i=75^\circ$. Image reproduced with permission from \cite{Schnittman09}, copyright by AAS.}
\label{fig:disk_spectrum}       
\end{figure}

Naturally, all of these relativistic effects are most prominent close to the black hole, where the gas is also the hottest and the emitted photons have the highest energies. Thus, there should be an apparent energy dependence to them. If we integrate over the entire disk, we can calculate the net intensity, polarization degree, and polarization angle, each as a function of energy, as are plotted in Figure~\ref{fig:disk_spectrum} for an $a/M= 0.998$, $L=0.1 L_\mathrm{Edd}$ system seen at an inclination of $i= 75^\circ$ from \cite{Schnittman09}. Each panel includes the total signal (solid curves), as well as the contributions from direct (dotted curves) and returning (dashed curves) radiation. 

Beginning with the spectral intensity (left panel of Figure~\ref{fig:disk_spectrum}), we see a broad thermal peak characteristic of a thin accretion disk. For black hole masses around $10 M_\odot$ and luminosities of $\sim 0.1 L_\mathrm{Edd}$, this thermal spectrum peaks around 1 keV. For a Novikov–Thorne (NT) emission profile, the direct radiation from a nonspinning black hole dominates over the returning radiation by a factor of roughly 100, with peak emissivity coming from around $R \simeq 9 M$. As the spin increases, the peak emission region moves closer to the horizon, and the relative fraction of observed flux that comes from returning radiation increases to $\sim 5$\% for $a/M= 0.9$ and $\sim 20$\% for $a/M= 0.998$ \cite{Schnittman09}. Because returning radiation is most prominent close to the black hole where the highest energy photons originate, the relative contribution from returning radiation increases with energy. From the left panel of Figure~\ref{fig:disk_spectrum}, we can see that for an $a/M= 0.998$ black hole, the returning radiation actually dominates the spectrum above $\sim 10$ keV.

Turning now to the polarization degree and angle in the middle and right panels of Figure~\ref{fig:disk_spectrum}, respectively, we see that at low energies (below the thermal peak), where relativistic effects are negligible, we get horizontal polarization ($\psi = 0^\circ$ or $180^\circ$) of a few percent, as expected \cite{Chandrasekhar60}. As a reminder, the degree of polarization at low energies is sensitive to inclination, increasing from $\delta \approx 1$\% for $i=45^\circ$ (not shown) to $\delta \approx 4$\% for $i = 75^\circ$. 

As the relative flux from returning radiation increases with increasing energy, we see the total polarization make a transition from the expected behavior of direct radiation to that of returning radiation. In the process of making this transition from horizontal polarization at low energies to vertical polarization at high energies, the degree of polarization goes through a local minimum as the two contributions cancel each other. For the $a/M=0.998$, $i=75^\circ$ case, this minimum occurs near 2 keV (middle panel). The polarization angle at that energy also begins to swing from $180^\circ$ to $90^\circ$ (right panel).

\section{Corona}
\label{sec:corona}

The previous section demonstrated that, while the polarization information coming from a thin disk is subject to a number of relativistic effects, it is generally possible to distinguish those effects, and potentially even use them to infer things about the system such as the black hole spin or disk inclination. In this section, we add our first potential complication to that picture by considering the role of the corona, the name given to the presumed cloud of hot electrons that give accreting black holes their hard X-ray tails through inverse Compton scattering of disk photons. One of the unresolved issues in high-energy astrophysics that polarization measurements may be able to help resolve is the physical geometry and location of this corona. Some numerical simulations have focused on trying to produce realistic coronae \citep{Kinch20}, but so far only one of them has looked at polarization signatures as a way to diagnose them \citep{Moscibrodzka2024}, which we look at in Section \ref{sec:states}. For now, as in the previous section, we consider a simplified analytic model. As one example, we present the so-called sandwich (or slab) corona, where the corona is distributed in a layer that lies on the top and bottom of the disk. One benefit of this approach is that it will simplify comparisons with the previous (disk-only) section. Alternative (though still idealized) geometries are considered in \cite{Schnittman10}.

Coronal scattering has two major effects on the observed properties of the disk. First, it hardens the energy spectrum by boosting some photons to higher energies and distributing them into a power law. Second, it changes the amplitude and orientation of the net polarization, especially at high energies. The impact on polarization, though, depends strongly on the size and location of the corona, as that will dictate which disk photons encounter it.

\begin{figure}
\sidecaption
\includegraphics[width=1.0\linewidth,trim=0mm 0mm 0mm 0,clip]{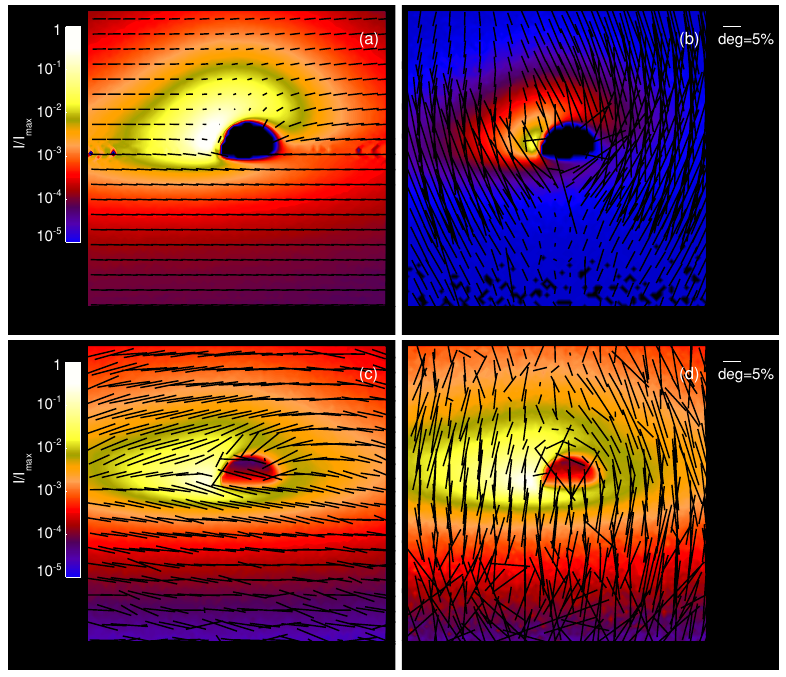}
%
%
\caption{Similar to Fig. \ref{fig:disk_image}, except for an accretion disk with a sandwich corona of scale height $H/R= 0.1$ and black hole spin of $a/M = 0.9$ (otherwise, same luminosity and inclination). The four panels correspond to (a) radiation coming directly from the thermal disk; (b) return radiation scattered once off the disk; (c) photons scattered once in the corona; and (d) photons scattered multiple times in the corona. Image reproduced with permission from \cite{Schnittman10}, copyright by AAS.}
\label{fig:corona_image}       
\end{figure}

Starting with the images of the disk + sandwich corona in Figure~\ref{fig:corona_image} and comparing them with the disk-only images of the previous section (Figure~\ref{fig:disk_image}), we can see subtle difference. Because the brightness of any given patch of the disk is dependent upon the optical depth along the geodesic connecting that patch with the observer, the presence of a corona can change the total optical depth along the path, but not equally for all paths, as it depends on the corona size, location, and geometry. For the sandwich geometry, in regions where the effective inclination is smallest (hence, shortest path through the corona), the observed flux is least affected, while in regions of high inclination (with longer paths through the corona), there will be additional scattering (greater limb darkening). The net effect to an image is a clockwise shift in the peak intensity distribution \citep[compare the left panel of Figure~\ref{fig:disk_image} with the upper-left image of Figure~\ref{fig:corona_image};][]{Schnittman10}. 

For return radiation, the main effect of a (sandwich) corona is to strongly suppress it for large radii. This is because, for return radiation to occur, the photons must pass close to the photon orbit ($M \le r_\mathrm{ph} \le 4M$, depending on the spin of the black hole), which necessarily means they encounter the outer disk at large angles of incidence. This means they face large optical depths through the corona. This suppression can be seen in the upper-right panel of Figure~\ref{fig:corona_image}. 

For high inclinations, photons will almost certainly scatter at least once in a sandwich corona before reaching an observer. If a photon scatters out mostly perpendicular to the sandwich corona, then the polarization will lie roughly parallel to the disk as in the outer parts of the image in the lower-left panel of Figure~\ref{fig:corona_image}. However, in cases where the photon undergoes multiple scatterings before leaving a sandwich corona, the scatterings will mostly happen within the disk plane, leading to a polarization angle perpendicular to it. This is similar to the case of returning radiation and explains why most of the polarization ticks in the bottom-right (multiple-scattering) panel of Figure~\ref{fig:corona_image} point vertically.

\begin{figure}
\sidecaption
\includegraphics[width=1.0\linewidth,trim=0mm 0mm 0mm 0,clip]{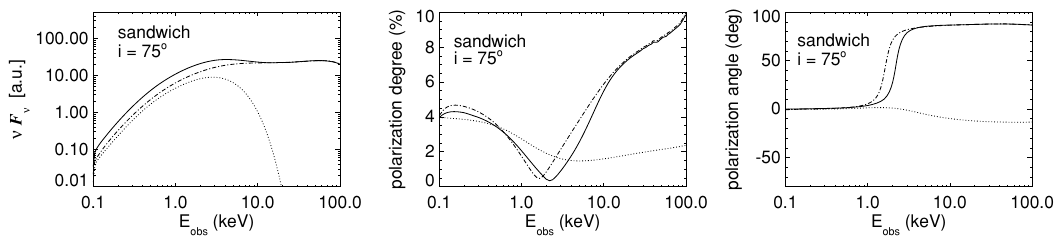}
%
%
\caption{Similar to Fig. \ref{fig:disk_spectrum}, except for an accretion disk with a sandwich corona of scale height $H/R= 0.1$ and black hole spin of $a/M = 0.9$ (otherwise, same luminosity and inclination). Each panel includes the direct (dotted curves), corona-scattered (dot-dashed curves), and total (solid curves) contributions. Image reproduced with permission from \cite{Schnittman10}, copyright by AAS.}
\label{fig:corona_spectrum}       
\end{figure}

As already mentioned, the main effect of a corona on the accretion disk spectrum is to introduce a hard tail that extends to higher energies than the disk's thermal peak. This tail can be seen in the left panel of Figure~\ref{fig:corona_spectrum} from \cite{Schnittman10}. The effects on the polarization degree and angle are mostly to exaggerate the same effects that were see in the disk-only case in the previous section. At low energies, where the outer disk dominates, the polarization is fairly low ($\lesssim$ a few percent) with $\psi \approx 0^\circ$ (or $180^\circ$). At high energies, large-angle scattering in the disk plane dominates, leading to higher polarization degrees ($\sim 10$\%) with $\psi \approx 90^\circ$. The transition happens near the thermal peak, where the polarization degree dips down toward 0\% (middle panel of Figure~\ref{fig:corona_spectrum}), as the two polarizations roughly cancel out, and the polarization angle swings from $0^\circ$ to $90^\circ$ \citep[right panel of Figure~\ref{fig:corona_spectrum};][]{Schnittman10, Tamborra18}.

In addition to the thermal inverse-Compton processes we have been focusing on so far, the photon energies can also be shifted by ``bulk Comptonization'' if the corona is moving rapidly relative to the disk \citep{Schnittman13a}. This is unlikely in a sandwich corona, but might be relevant if the corona were part of the jet, which leads us nicely into the next section of our chapter.

\section{Jet}
\label{sec:jet}

In the previous two sections, we focused on simplified models of accretion disks and coronae to help disentangle the many effects of relativity. However, our choice was also partly motivated by the fact that there are still very few GRMHD simulations focused on accretion disks that report polarization measurements. Luckily, there are many more GRMHD simulations focused on jets that have reported polarization signatures \cite{Broderick10, Shcherbakov12, Gold17, Moscibrodzka17, Tsunetoe20, Davelaar23}, though mostly in the context of active galactic nuclei (AGN) and radio (rather than X-ray) polarimetry. Nevertheless, this gives us an opportunity to discuss important new effects, such as synchrotron radiation and Faraday rotation.

The (radio) jets associated with black hole accretion disks are thought to be powered by the rotational energy of the black hole \citep{Blandford77} or accretion disk \citep{Blandford82}, with strong magnetic fields acting as the conduit to carry this energy away. Provided there is a supply of relativistic electrons, these strong magnetic fields can cause the electrons to produce synchrotron radiation, which is intrinsically highly linearly polarized with the polarization vector perpendicular to the magnetic field lines \cite{Rybicki79}. Thus, at least in some cases, the dominant source of X-ray polarization could be the jet.

\begin{figure}
\sidecaption
\includegraphics[width=1.0\linewidth,trim=0mm 0mm 0mm 0,clip]{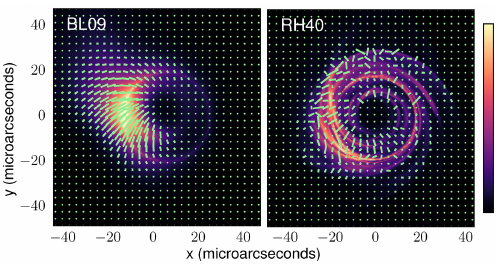}
%
%
\caption{Intensity (colors) and polarization maps (ticks) comparing a semi-analytic force-free jet model \citep[left;][]{Broderick09} with a GRMHD simulation result \citep[right;][]{Moscibrodzka17}. In both cases, the observer viewing angle is fixed to $i= 20^\circ$ from the black hole spin axis. In the semi-analytic model, radiation from the approaching foreground jet shows coherent polarization (net polarization of $\approx 15$\%) that traces the helical magnetic field structure. In the GRMHD simulation, radiation from the receding background jet undergoes strong Faraday rotation as it passes through the accretion disk, leading to the relatively scrambled polarization pattern (net polarization of $\approx 1$\%). Image reproduced with permission from \cite{Moscibrodzka17}, copyright by RAS.}
\label{fig:jet_Faraday}       
\end{figure}

However, even in AGN with strong jets, the observed (radio) linear polarization is generally quite low \cite{Zavala03, Park19}. This is likely due to Faraday rotation, which occurs when polarized light travels through a magnetized plasma, affecting the propagation speed and shifting the polarization plane by an amount proportional to the rotation measure (RM)\footnote{RM is constant for an external Faraday screen. However, when the Faraday screen is internal to the emission zone, RM can additionally depend on the observing wavelength through opacity effects, see for example \cite{Hovatta2019}.}:
\begin{equation}
\mathrm{RM} = \frac{\psi(\lambda_1)-\psi(\lambda_2)}{\lambda_1^2 - \lambda_2^2} = \frac{e^3}{2\pi m_\mathrm{e}^2 c^4} \int n_e B_\parallel \mathrm{d}s \,[\mathrm{rad~cm^{-2}}] ~.
\end{equation}
Figure~\ref{fig:jet_Faraday} from \cite{Moscibrodzka17} provides an example of the depolarization associated with a foreground Faraday screen, showing how different emission zones can experience different amounts of Faraday rotation. Whenever there is no Faraday rotation (left panel), the polarization degree is quite high ($\gtrsim 15$\%) and well organized, revealing the structure of the helical magnetic field in the jet. On the other hand, whenever there is a Faraday screen in the way (right panel), the net polarization degree drops considerably ($\lesssim 1$\%) and becomes quite chaotic. 

\begin{figure}
\sidecaption
\includegraphics[width=1.0\linewidth,trim=0mm 0mm 0mm 0,clip]{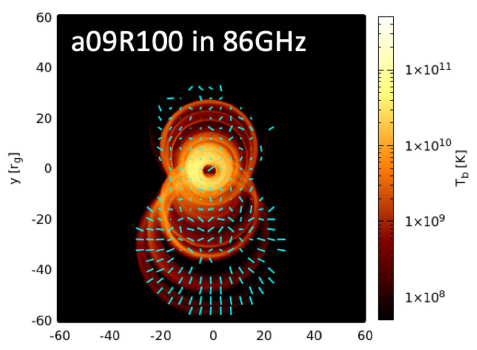}
%
%
\caption{An 86 GHz intensity map overlaid with linear polarization vectors from a GRMHD simulation of the M87* jet. The polarization is much stronger and more organized for the approaching jet (bottom of image) than for the receding (top). Image reproduced with permission from \cite{Tsunetoe20}, copyright by ASJ.}
\label{fig:jet_image}       
\end{figure}

Another example is shown in Figure~\ref{fig:jet_image} from \cite{Tsunetoe20}, where we observe both the approaching (in the lower part) and receding (in the upper part) jets from a simulation designed to model M87*. We can see that both jets show limb-brightening due to the high Lorentz factors in the jet rims. We also see a strong difference in the linear polarization, with the receding jet showing much weaker polarization, as its light must pass through the accretion disk, which acts as an effective Faraday screen. Thus, we see that polarization measurements of synchrotron jets can provide detailed information about the relative orientation and structure of magnetic fields in the vicinity of black holes, although Faraday rotation can be a significant hindrance by reducing the overall polarization. 

\section{Black Hole Spectral States}
\label{sec:states}

Many black hole (and neutron star) X-ray binaries (XRBs) transit between different ``states'' during their outbursts. Primarily, the states are identified by their luminosities and the presence or absence of different components in the spectra \cite{Belloni10}. The two major spectral states, accompanied by various intermediate ones, are the ``soft'' and ``hard'' states. In the soft state, the spectra are predominantly composed of a soft X-ray component (usually peaking below 3 keV). This soft emission likely arises from the disk, exhibiting thermal blackbody-like radiation. On the other hand, in the hard state, the spectra are dominated by a power-law component (extending roughly between 10 - 100 keV). This power-law emission likely results from the Compton upscattering of soft seed photons emitted from the disk by the corona. 

A useful way to identify these states and the transitions between them during an outburst is by tracing the movement of the source through a hardness-intensity diagram  \cite{McClintock06}. In the case of black hole XRBs, they often trace out a characteristic ``q'' pattern in such diagrams \cite{Fender04}. Black hole XRBs also often exhibit characteristic types and levels of variability associated with each state \cite{Ingram19}, which we will touch on in the next section. New work is trying to see how polarization varies across these states \citep{Majumder26}.

The location and geometry of the corona are still the subject of considerable debate; some models suggest that it resides at the base of the jet \citep{Martocchia96, Fabian09}, while others consider it to sandwich the disk \citep{Galeev79, Haardt93a, Beloborodov99}, as presented in Section \ref{sec:corona}. Alternatively, the truncated-disk model envisions a third possible geometry with a standard, thin disk truncated well outside the last stable orbit, and the corona as a geometrically thick, radiatively inefficient accretion flow filling the inner gap \citep{Eardley75, Esin97}. Despite the uncertainty regarding the location of the corona, all of these models can reproduce the observed X-ray spectra during the state transitions by associating them with appropriate changes in the geometry and strength of the accretion disk and corona, as well as the presence or absence of other physical components such as jets. 

\begin{figure}
\sidecaption
\includegraphics[width=1.0\linewidth,trim=0mm 0mm 0mm 0,clip]{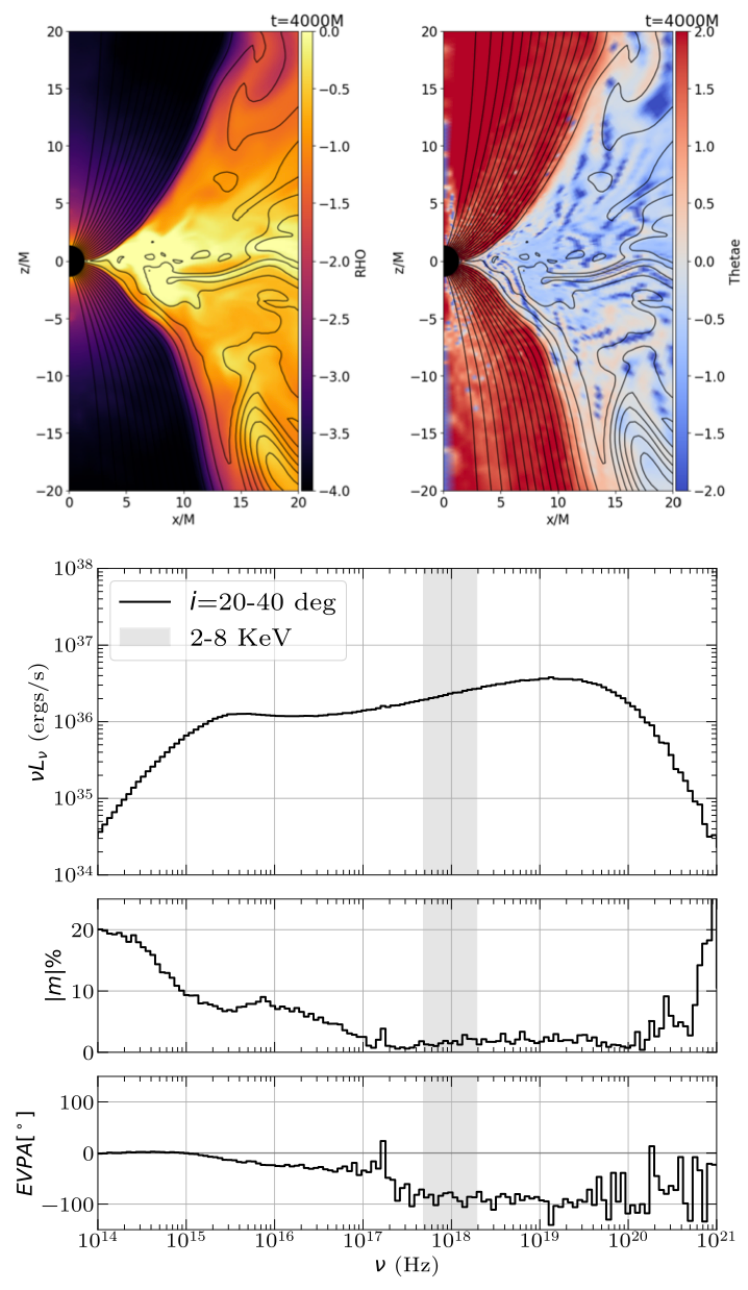}
\caption{Slices through a GRMHD simulation intended to represent the hard state of a black hole XRB showing the {\em Top left:} plasma density and {\em Top right:} dimensionless electron temperature. {\em Bottom:} Spectra of total intensity, polarization degree, and polarization angle. The gray shaded region marks the IXPE band (2-8 keV). Image reproduced with permission from \cite{Moscibrodzka2024}.}
\label{fig:CygX1}       
\end{figure}

Polarimetry combined with spectroscopic and timing analyses has the potential to break this degeneracy \cite{Haardt93b}. A first attempt to apply GRRMHD simulations + polarized radiative transport toward this goal was presented in \cite{Moscibrodzka2024}.  Figure~\ref{fig:CygX1} shows slices from a simulation meant to represent the hard state of a black hole XRB, along with spectra of the intensity, polarization degree, and polarization angle for one particular observer inclination. More work needs to be done, though, to test other hard-state geometries and other accretion states. Nevertheless, this demonstrates the potential of coupling GRMHD simulations to polarized radiation transport as a way to interpret observations by IXPE or other future X-ray polarimeters.

\section{Time Variability}
\label{sec:variability}

XRBs also exhibit interesting temporal variability in their X-ray light curves, often in the form of quasi-periodic oscillations (QPOs) \citep{Remillard06, Belloni10}. For black hole systems, these QPOs are classified as high- ($\gtrsim 60$ Hz) and low-frequency ($\lesssim 30$ Hz), with the low-frequency, especially the type-C, having properties strongly suggesting a geometric origin \citep{Fragile23b}. A promising model for this QPO is having a finite, inner region of the accretion flow precess as a rigid body \citep{Ingram09}. Thus, the time variability and geometry of black hole accretion flows may be closely linked, and polarization measurements may be able to shed light on both.

\begin{figure}
\sidecaption
\includegraphics[width=1.0\linewidth,trim=0mm 0mm 0mm 0,clip]{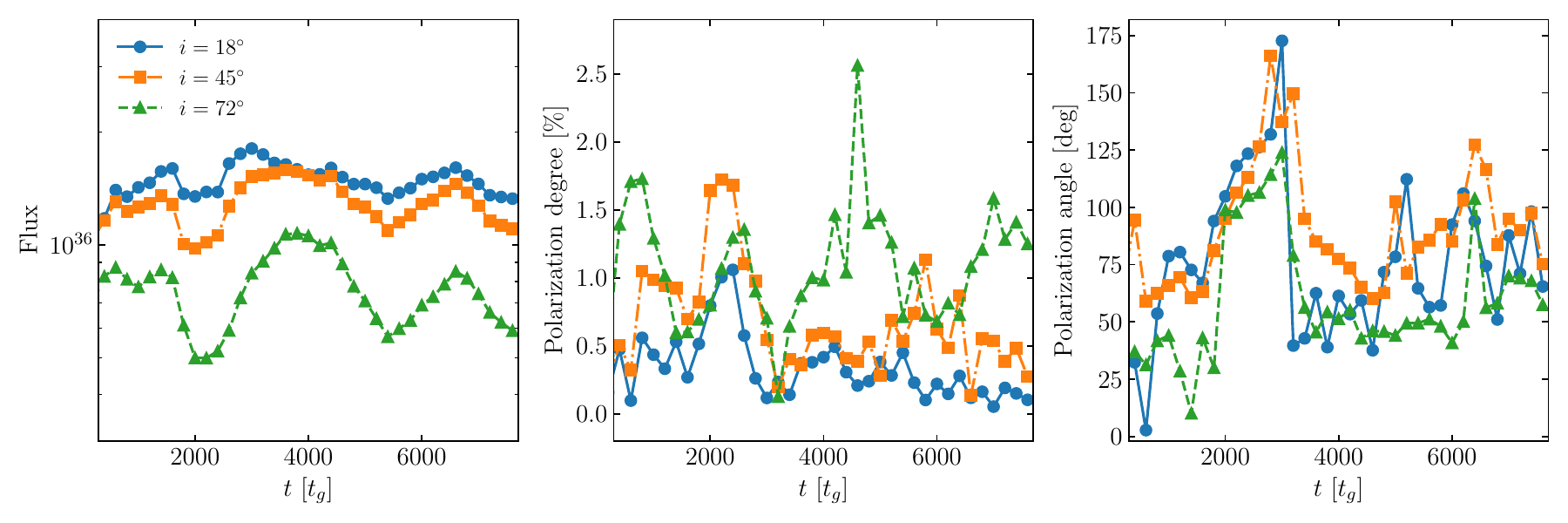}
%
%
\caption{{\em Left:} Intensity, {\em Middle:} polarization degree, and {\em Right:} polarization angle plotted as a function of time for three different observer inclinations integrated over the IXPE sensitivity range for an isolated tilted precessing torus. This plot covers about 1.7 precession periods, which are most visible in the intensity and polarization angle plots. Image reproduced with permission from \cite{Fragile25b}, copyright by AAS.}
\label{fig:tilted_torus}       
\end{figure}

Recent simulations have been designed to study tilted, precessing accretion disks \citep{Bollimpalli24}, and the results have now been processed through {\tt Pandurata} to produce polarization maps, spectra, and light curves \citep{Fragile25b}. To illustrate one case, in Figure~\ref{fig:tilted_torus} we plot the intensity, polarization degree, and polarization angle as a function of time for a tilted, isolated torus. This case exhibits prominent fluctuations in the flux and polarization measures roughly on the precession period ($4500\, GM/c^3$). The polarization angle, in particular, clearly traces the first precession period for all observer inclinations with an amplitude $\gtrsim 30^\circ$. The polarization degree and the late-time polarization angle are less well synchronized with the precession, though the intermediate inclination ($i=45^\circ$) does show roughly two periods of oscillation. These results confirm that polarization can be a powerful diagnostic for testing models of time variability in black hole accretion disks, especially if they have a geometric origin.

\section{Event Horizon Telescope}
\label{sec:EHT}

Having demonstrated the general potential of coupling GRMHD simulations with polarized radiative transport, we now want to discuss one case where it is already being put into action -- in interpreting observations of the Event Horizon Telescope (EHT). There are two caveats that we should get out of the way beforehand. First, EHT is largely focused on imaging on the scale of individual black holes, which, as we mentioned in Section \ref{sec:transport}, is not possible with an X-ray instrument. Second, the polarized emission in EHT sources comes from synchrotron radiation of electrons gyrating around magnetic field lines, which is distinct from the scattering-dominated origin of polarization in the context of X-ray binaries. Nevertheless, the EHT studies constitute the most thorough and mature applications of GRMHD polarimetry to the theoretical interpretation of astronomical observations, and as such could inform the future developments of similar analyses for X-ray polarimetry. 

Observing images of accreting black hole systems on horizon scales offers an unprecedented opportunity to test numerical GRMHD models of accretion. 
EHT observations give access to the resolved morphology of the compact emission region and include complete polarimetric information, incorporating both linear and circular polarization components \citep{EHTC_M87_p8,EHTC_M87_p9,EHTC_SgrA_p8}.

\begin{figure}
\sidecaption
\includegraphics[width=1.0\linewidth,trim=0mm 0mm 0mm 0,clip]{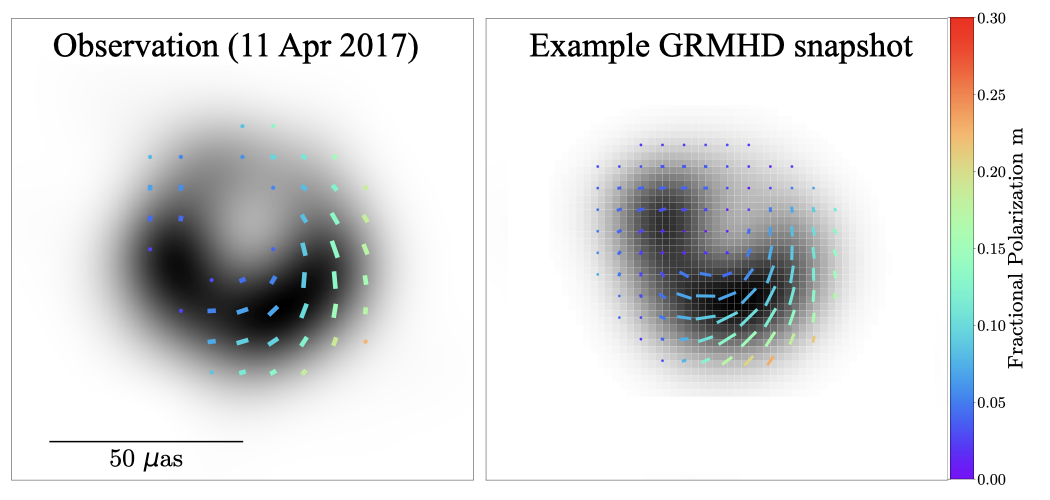}
%
%
\caption{{\em Left:} Observation of M87* by the EHT. Dark heatmap in the background represents the total intensity image morphology. Ticks represent the polarized emission, with their orientation representing the polarization angle, length proportional to the polarized flux density, and color related to the polarization degree. {\em Right:} A ray-traced snapshot from a GRMHD simulation of M87*, assuming a strongly magnetized accretion flow around a black hole with spin $a_* = 0.5$. Image was blurred to the approximate instrumental resolution of the EHT. Image adapted with permission from \cite{EHTC_M87_p8}, copyright by AAS.}
\label{fig:obs_GRMHD}       
\end{figure}

The general approach of the EHT to these analyses is to construct large libraries of synthetic GRMHD images computed through ray tracing, extract particular observable parameters, and characterize their statistical consistency with the observations (the simulations are not strictly stationary as the modeled accretion flows are turbulent). An example snapshot image from the M87* library is shown in the right panel of Figure~\ref{fig:obs_GRMHD}, with the EHT observation presented on the left. The libraries, composed of $\sim 10^5$ snapshot images of M87* \cite{EHTC_M87_p5,EHTC_M87_p8,EHTC_M87_p9} and $\sim 10^6$ snapshot images of Sgr~A* \cite{EHTC_SgrA_p5,EHTC_SgrA_p8}, systematically explore a number of theoretical parameters such as:
\begin{itemize}
    \item black hole spin $a$ (GRMHD setup parameter),
    \item strongly or weakly magnetized state of the accretion (GRMHD setup parameter),
    \item electron temperature model (radiation transport postprocessing parameters following \cite{Moscibrodzka2016}, important as a plasma physics constraint),
    \item viewing angle of the source (radiation transport postprocessing parameter, poorly constrained for Sgr~A* but informed by the observed jet in the case of M87*),
    \item magnetic field polarity (simulation postprocessing parameter, particularly relevant for the circular polarization \cite{EHTC_M87_p9,EHTC_SgrA_p8, Joshi2024} ).
\end{itemize}
The plasma density in these simulations is scaled to match the observed radiative flux density measured by the EHT under independent mass and distance priors. Additional sets of synthetic images explore issues such as GRMHD and ray-tracing code consistency \cite{Porth19,Prather2023}, non-thermal distribution of the energy of the electrons \citep{Davelaar2019}, tilt of the accretion disk \citep{Liska19}, inclusion of radiation within the GRMHD simulation \cite{Chael2019}, or deviation from Kerr geometry \cite{EHTC_SgrA_p6}.

Polarimetric constraints used by the EHT include both structural (resolved) ones, such as the image-averaged absolute value of the polarization fraction or coefficients of the azimuthal Fourier decomposition of the linear polarization structure, and unresolved ones, such as the polarization vector averaged over the entire image. The latter type include the net degree of linear polarization $\delta$ as well as the net polarization angle $\psi$, quantities that are equivalent to those used in X-ray domain studies on spatially unresolved sources.

For both horizon-scale sources, GRMHD-aided polarimetric analysis by the EHT has demonstrated the existence of coherent magnetic fields in the immediate environment of the event horizon. Comparisons with numerical simulations indicate the presence of a dynamically important poloidal magnetic field component, favoring strongly magnetized theoretical models over weakly magnetized ones that generally indicate much larger Faraday depths for both EHT sources \citep{EHTC_M87_p8,EHTC_SgrA_p8}. Furthermore, the depolarizing impact of Faraday rotation, discussed in Section \ref{sec:jet}, has been demonstrated \citep{EHTC_M87_p8}. Analysis favors a low inclination for the accretion disk in Sgr~A* (\citep{EHTC_SgrA_p5, EHTC_M87_p8}; while perhaps a bit surprising based on our viewing location within the Galaxy, this conclusion is consistent with the independent analyses of \cite{Gravity2018, Wielgus2022polar}). The asymmetry of the resolved M87* image disfavors low black hole spin values \citep{EHTC_M87_p5, Saurabh2025}. Models with strong thermal coupling between ions and electrons are effectively ruled out, confirming observationally the expected two-temperature character of low-mass-accretion-rate, advection-dominated flows \cite{Yuan2014}. In some cases, polarimetric constraints restrict the parameter space of numerical models much more than the total intensity constraints alone, resulting in improved estimates of certain physical properties such as the radiative efficiency and mass accretion rate \citep{EHTC_M87_p8}. Recent EHT results comparing the total intensity and linear polarization of M87* in 2017, 2018, and 2021 demonstrate the time dependence of the system, particularly the evolution of the Faraday screen \cite{EHT_M87_2025}. 

The GRMHD models considered by the EHT have succeeded in capturing the observed averaged properties of M87* and Sgr~A* to a large degree. However, there seems to be a persistent discrepancy between the observed horizon-scale dynamics and simulations, with the latter overproducing the amount of variability of both total intensity \citep{EHTC_SgrA_p5, Wielgus2022EHT} and polarimetric observables \cite{Wielgus2024}. This disagreement is already driving new developments in numerical GRMHD \citep{Salas2025,Dhruv2025}.

Finally, the EHT has delivered extreme-resolution polarimetric observations for a number of AGN sources, resolving the structure of their compact jets \cite{Issaoun2022,Jorstad2023,Paraschos2024, Roeder2025}. While quantitative comparisons with GRMHD at the relevant spatial scales has been limited so far, there are very interesting future prospects of combining multiwavelength analysis of radio and X-ray polarimetric observations \cite{Chen2024} with theoretical interpretations from numerical simulations.

\section{Final Thoughts \& Future Directions}
\label{sec:conclusion}

At the start of this chapter, we listed some of the many GRMHD and GRRMHD simulations of black hole accretion that have been published to date. The point in doing so was to emphasize that {\em all} of those simulations are potential inputs to polarized radiation post-processing codes like the ones described in Section \ref{sec:transport}. We are still very early in the era of X-ray polarimetry, so it is going to be useful to have a wide array of predictions for what we might expect.

Here is our list of things that we feel could and should be done in the near future (not in any particular order):
\begin{enumerate}
\item One of the big open questions in black hole accretion physics is the location and geometry of the corona, which is an especially important component in the hard spectral state. It will, therefore, be helpful for simulations that claim to represent that state to be post-processed for their polarization signatures to compare with observations. This may help us estimate the validity of different coronal models.
\item Relatively little has been done to measure the effects of Faraday rotation from simulations. Some work has been done for shearing box simulations \cite{Davis09} and in the context of EHT observations \cite{Moscibrodzka17, Tsunetoe20}, but nothing has been done for X-ray polarization from global simulations. There are claims that the lack of Faraday rotation seen in IXPE observations places tight constraints on the strength of the magnetic fields in accreting black hole systems \cite{Barnier24}. This certainly needs to be tested, and simulations provide a straightforward way to do so.
\item Some simulations are beginning to produce QPO-like features, which may be testable based on their polarization signatures \cite{Fragile25b}. Although IXPE has had limited success in finding polarization changes associated with QPOs \citep{Zhao24}, this should become more possible with future instruments, so this is another avenue worth pursuing.
\end{enumerate}

\begin{acknowledgement}
PCF gratefully acknowledges the support of NASA through award No 23-ATP23-0100. MW is supported by a Ramón y Cajal grant RYC2023-042988-I from the Spanish Ministry of Science and Innovation and acknowledges financial support from the Severo Ochoa grant CEX2021-001131-S funded by MCIN/AEI/ 10.13039/501100011033. The work also benefits from collaboration supported by the Czech Grant Agency under project no. 21-06825X, ``Accretionary black holes in the new era of X-ray polarimetry.'' The Flatiron Institute is a division of the Simons Foundation.

\end{acknowledgement}

\bibliographystyle{utphys}

\begin{thebibliography}{100}

\bibitem{Shafee08}
R.~{Shafee}, J.~C. {McKinney}, R.~{Narayan}, A.~{Tchekhovskoy}, C.~F. {Gammie},
  and J.~E. {McClintock}, ``{Three-Dimensional Simulations of Magnetized Thin
  Accretion Disks around Black Holes: Stress in the Plunging Region},''
  \href{https://dx.doi.org/10.1086/593148}{{\em \apjl} {\bfseries 687} no.~1,
  (Nov., 2008) L25}, \href{https://arxiv.org/abs/0808.2860}{{\ttfamily
  arXiv:0808.2860 [astro-ph]}}.

\bibitem{Penna10}
R.~F. {Penna}, J.~C. {McKinney}, R.~{Narayan}, A.~{Tchekhovskoy}, R.~{Shafee},
  and J.~E. {McClintock}, ``{Simulations of magnetized discs around black
  holes: effects of black hole spin, disc thickness and magnetic field
  geometry},'' \href{https://dx.doi.org/10.1111/j.1365-2966.2010.17170.x}{{\em
  \mnras} {\bfseries 408} no.~2, (Oct., 2010) 752--782},
  \href{https://arxiv.org/abs/1003.0966}{{\ttfamily arXiv:1003.0966
  [astro-ph.HE]}}.

\bibitem{Penna12}
R.~F. {Penna}, A.~{S{\k{a}}dowski}, and J.~C. {McKinney}, ``{Thin-disc theory
  with a non-zero-torque boundary condition and comparisons with
  simulations},''
  \href{https://dx.doi.org/10.1111/j.1365-2966.2011.20084.x}{{\em \mnras}
  {\bfseries 420} no.~1, (Feb., 2012) 684--698},
  \href{https://arxiv.org/abs/1110.6556}{{\ttfamily arXiv:1110.6556
  [astro-ph.HE]}}.

\bibitem{Sadowski16}
A.~{S{\k{a}}dowski}, ``{Thin accretion discs are stabilized by a strong
  magnetic field},'' \href{https://dx.doi.org/10.1093/mnras/stw913}{{\em
  \mnras} {\bfseries 459} no.~4, (July, 2016) 4397--4407},
  \href{https://arxiv.org/abs/1601.06785}{{\ttfamily arXiv:1601.06785
  [astro-ph.HE]}}.

\bibitem{Lancova2019}
D.~{Lan{\v{c}}ov{\'a}}, D.~{Abarca}, W.~{Klu{\'z}niak}, M.~{Wielgus},
  A.~{S{\k{a}}dowski}, R.~{Narayan}, J.~{Schee}, G.~{T{\"o}r{\"o}k}, and
  M.~{Abramowicz}, ``{Puffy Accretion Disks: Sub-Eddington, Optically Thick,
  and Stable},'' \href{https://dx.doi.org/10.3847/2041-8213/ab48f5}{{\em \apjl}
  {\bfseries 884} no.~2, (Oct., 2019) L37},
  \href{https://arxiv.org/abs/1908.08396}{{\ttfamily arXiv:1908.08396
  [astro-ph.HE]}}.

\bibitem{Mishra22}
B.~{Mishra}, P.~C. {Fragile}, J.~{Anderson}, A.~{Blankenship}, H.~{Li}, and
  K.~{Nalewajko}, ``{The Role of Strong Magnetic Fields in Stabilizing Highly
  Luminous Thin Disks},''
  \href{https://dx.doi.org/10.3847/1538-4357/ac938b}{{\em \apj} {\bfseries 939}
  no.~1, (Nov., 2022) 31}, \href{https://arxiv.org/abs/2209.03317}{{\ttfamily
  arXiv:2209.03317 [astro-ph.HE]}}.

\bibitem{DeVilliers03}
J.-P. {De Villiers} and J.~F. {Hawley}, ``{Global General Relativistic
  Magnetohydrodynamic Simulations of Accretion Tori},''
  \href{https://dx.doi.org/10.1086/375866}{{\em \apj} {\bfseries 592} no.~2,
  (Aug., 2003) 1060--1077},
  \href{https://arxiv.org/abs/astro-ph/0303241}{{\ttfamily
  arXiv:astro-ph/0303241 [astro-ph]}}.

\bibitem{Gammie03}
C.~F. {Gammie}, J.~C. {McKinney}, and G.~{T{\'o}th}, ``{HARM: A Numerical
  Scheme for General Relativistic Magnetohydrodynamics},''
  \href{https://dx.doi.org/10.1086/374594}{{\em \apj} {\bfseries 589} no.~1,
  (May, 2003) 444--457},
  \href{https://arxiv.org/abs/astro-ph/0301509}{{\ttfamily
  arXiv:astro-ph/0301509 [astro-ph]}}.

\bibitem{Narayan12}
R.~{Narayan}, A.~{S{\k{a}}dowski}, R.~F. {Penna}, and A.~K. {Kulkarni},
  ``{GRMHD simulations of magnetized advection-dominated accretion on a
  non-spinning black hole: role of outflows},''
  \href{https://dx.doi.org/10.1111/j.1365-2966.2012.22002.x}{{\em \mnras}
  {\bfseries 426} no.~4, (Nov., 2012) 3241--3259},
  \href{https://arxiv.org/abs/1206.1213}{{\ttfamily arXiv:1206.1213
  [astro-ph.HE]}}.

\bibitem{Shiokawa12}
H.~{Shiokawa}, J.~C. {Dolence}, C.~F. {Gammie}, and S.~C. {Noble}, ``{Global
  General Relativistic Magnetohydrodynamic Simulations of Black Hole Accretion
  Flows: A Convergence Study},''
  \href{https://dx.doi.org/10.1088/0004-637X/744/2/187}{{\em \apj} {\bfseries
  744} no.~2, (Jan., 2012) 187},
  \href{https://arxiv.org/abs/1111.0396}{{\ttfamily arXiv:1111.0396
  [astro-ph.HE]}}.

\bibitem{Yuan12}
F.~{Yuan}, M.~{Wu}, and D.~{Bu}, ``{Numerical Simulation of Hot Accretion
  Flows. I. A Large Radial Dynamical Range and the Density Profile of Accretion
  Flow},'' \href{https://dx.doi.org/10.1088/0004-637X/761/2/129}{{\em \apj}
  {\bfseries 761} no.~2, (Dec., 2012) 129},
  \href{https://arxiv.org/abs/1206.4157}{{\ttfamily arXiv:1206.4157
  [astro-ph.HE]}}.

\bibitem{Sadowski13a}
A.~{S{\k{a}}dowski}, R.~{Narayan}, R.~{Penna}, and Y.~{Zhu}, ``{Energy,
  momentum and mass outflows and feedback from thick accretion discs around
  rotating black holes},'' \href{https://dx.doi.org/10.1093/mnras/stt1881}{{\em
  \mnras} {\bfseries 436} no.~4, (Dec., 2013) 3856--3874},
  \href{https://arxiv.org/abs/1307.1143}{{\ttfamily arXiv:1307.1143
  [astro-ph.HE]}}.

\bibitem{Dibi12}
S.~{Dibi}, S.~{Drappeau}, P.~C. {Fragile}, S.~{Markoff}, and J.~{Dexter},
  ``{General relativistic magnetohydrodynamic simulations of accretion on to
  Sgr A*: how important are radiative losses?},''
  \href{https://dx.doi.org/10.1111/j.1365-2966.2012.21857.x}{{\em \mnras}
  {\bfseries 426} no.~3, (Nov., 2012) 1928--1939},
  \href{https://arxiv.org/abs/1206.3976}{{\ttfamily arXiv:1206.3976
  [astro-ph.HE]}}.

\bibitem{Sadowski17}
A.~{S{\k{a}}dowski}, M.~{Wielgus}, R.~{Narayan}, D.~{Abarca}, J.~C. {McKinney},
  and A.~{Chael}, ``{Radiative, two-temperature simulations of low-luminosity
  black hole accretion flows in general relativity},''
  \href{https://dx.doi.org/10.1093/mnras/stw3116}{{\em \mnras} {\bfseries 466}
  no.~1, (Apr., 2017) 705--725},
  \href{https://arxiv.org/abs/1605.03184}{{\ttfamily arXiv:1605.03184
  [astro-ph.HE]}}.

\bibitem{Jiang19}
Y.-F. {Jiang}, O.~{Blaes}, J.~M. {Stone}, and S.~W. {Davis}, ``{Global
  Radiation Magnetohydrodynamic Simulations of sub-Eddington Accretion Disks
  around Supermassive Black Holes},''
  \href{https://dx.doi.org/10.3847/1538-4357/ab4a00}{{\em \apj} {\bfseries 885}
  no.~2, (Nov., 2019) 144}, \href{https://arxiv.org/abs/1904.01674}{{\ttfamily
  arXiv:1904.01674 [astro-ph.HE]}}.

\bibitem{Dexter21}
J.~{Dexter}, N.~{Scepi}, and M.~C. {Begelman}, ``{Radiation GRMHD Simulations
  of the Hard State of Black Hole X-Ray Binaries and the Collapse of a Hot
  Accretion Flow},'' \href{https://dx.doi.org/10.3847/2041-8213/ac2608}{{\em
  \apjl} {\bfseries 919} no.~2, (Oct., 2021) L20},
  \href{https://arxiv.org/abs/2109.06239}{{\ttfamily arXiv:2109.06239
  [astro-ph.HE]}}.

\bibitem{Jiang14}
Y.-F. {Jiang}, J.~M. {Stone}, and S.~W. {Davis}, ``{A Global Three-dimensional
  Radiation Magneto-hydrodynamic Simulation of Super-Eddington Accretion
  Disks},'' \href{https://dx.doi.org/10.1088/0004-637X/796/2/106}{{\em \apj}
  {\bfseries 796} no.~2, (Dec., 2014) 106},
  \href{https://arxiv.org/abs/1410.0678}{{\ttfamily arXiv:1410.0678
  [astro-ph.HE]}}.

\bibitem{Sadowski14}
A.~{S{\k{a}}dowski}, R.~{Narayan}, J.~C. {McKinney}, and A.~{Tchekhovskoy},
  ``{Numerical simulations of super-critical black hole accretion flows in
  general relativity},'' \href{https://dx.doi.org/10.1093/mnras/stt2479}{{\em
  \mnras} {\bfseries 439} no.~1, (Mar., 2014) 503--520},
  \href{https://arxiv.org/abs/1311.5900}{{\ttfamily arXiv:1311.5900
  [astro-ph.HE]}}.

\bibitem{McKinney14}
J.~C. {McKinney}, A.~{Tchekhovskoy}, A.~{Sadowski}, and R.~{Narayan},
  ``{Three-dimensional general relativistic radiation magnetohydrodynamical
  simulation of super-Eddington accretion, using a new code HARMRAD with M1
  closure},'' \href{https://dx.doi.org/10.1093/mnras/stu762}{{\em \mnras}
  {\bfseries 441} no.~4, (July, 2014) 3177--3208},
  \href{https://arxiv.org/abs/1312.6127}{{\ttfamily arXiv:1312.6127
  [astro-ph.CO]}}.

\bibitem{Utsumi22}
A.~{Utsumi}, K.~{Ohsuga}, H.~R. {Takahashi}, and Y.~{Asahina}, ``{Component of
  Energy Flow from Supercritical Accretion Disks Around Rotating Stellar Mass
  Black Holes},'' \href{https://dx.doi.org/10.3847/1538-4357/ac7eb8}{{\em \apj}
  {\bfseries 935} no.~1, (Aug., 2022) 26},
  \href{https://arxiv.org/abs/2207.02560}{{\ttfamily arXiv:2207.02560
  [astro-ph.HE]}}.

\bibitem{Yoshioka22}
S.~{Yoshioka}, S.~{Mineshige}, K.~{Ohsuga}, T.~{Kawashima}, and T.~{Kitaki},
  ``{Large-scale outflow structure and radiation properties of super-Eddington
  flow: Dependence on the accretion rates},''
  \href{https://dx.doi.org/10.1093/pasj/psac076}{{\em \pasj} {\bfseries 74}
  no.~6, (Dec., 2022) 1378--1395},
  \href{https://arxiv.org/abs/2209.01427}{{\ttfamily arXiv:2209.01427
  [astro-ph.HE]}}.

\bibitem{Fragile25}
P.~C. {Fragile}, M.~J. {Middleton}, D.~A. {Bollimpalli}, and Z.~{Smith},
  ``{Long time-scale numerical simulations of large supercritical accretion
  discs},'' \href{https://dx.doi.org/10.1093/mnras/staf890}{{\em \mnras}
  {\bfseries 540} no.~3, (July, 2025) 2820--2829},
  \href{https://arxiv.org/abs/2505.08859}{{\ttfamily arXiv:2505.08859
  [astro-ph.HE]}}.

\bibitem{Zhang25}
L.~{Zhang}, J.~M. {Stone}, P.~D. {Mullen}, S.~W. {Davis}, Y.-F. {Jiang}, and
  C.~J. {White}, ``{Radiation GRMHD Models of Accretion onto Stellar-Mass Black
  Holes: I. Survey of Eddington Ratios},''
  \href{https://dx.doi.org/10.48550/arXiv.2506.02289}{{\em arXiv e-prints}
  (June, 2025) arXiv:2506.02289},
  \href{https://arxiv.org/abs/2506.02289}{{\ttfamily arXiv:2506.02289
  [astro-ph.HE]}}.

\bibitem{Beckwith08}
K.~{Beckwith}, J.~F. {Hawley}, and J.~H. {Krolik}, ``{The Influence of Magnetic
  Field Geometry on the Evolution of Black Hole Accretion Flows: Similar Disks,
  Drastically Different Jets},'' \href{https://dx.doi.org/10.1086/533492}{{\em
  \apj} {\bfseries 678} no.~2, (May, 2008) 1180--1199},
  \href{https://arxiv.org/abs/0709.3833}{{\ttfamily arXiv:0709.3833
  [astro-ph]}}.

\bibitem{Tchekhovskoy11}
A.~{Tchekhovskoy}, R.~{Narayan}, and J.~C. {McKinney}, ``{Efficient generation
  of jets from magnetically arrested accretion on a rapidly spinning black
  hole},'' \href{https://dx.doi.org/10.1111/j.1745-3933.2011.01147.x}{{\em
  \mnras} {\bfseries 418} no.~1, (Nov., 2011) L79--L83},
  \href{https://arxiv.org/abs/1108.0412}{{\ttfamily arXiv:1108.0412
  [astro-ph.HE]}}.

\bibitem{McKinney12}
J.~C. {McKinney}, A.~{Tchekhovskoy}, and R.~D. {Blandford}, ``{General
  relativistic magnetohydrodynamic simulations of magnetically choked accretion
  flows around black holes},''
  \href{https://dx.doi.org/10.1111/j.1365-2966.2012.21074.x}{{\em \mnras}
  {\bfseries 423} no.~4, (July, 2012) 3083--3117},
  \href{https://arxiv.org/abs/1201.4163}{{\ttfamily arXiv:1201.4163
  [astro-ph.HE]}}.

\bibitem{Wielgus2015}
M.~{Wielgus}, P.~C. {Fragile}, Z.~{Wang}, and J.~{Wilson}, ``{Local stability
  of strongly magnetized black hole tori},''
  \href{https://dx.doi.org/10.1093/mnras/stu2676}{{\em \mnras} {\bfseries 447}
  no.~4, (Mar., 2015) 3593--3601},
  \href{https://arxiv.org/abs/1412.4561}{{\ttfamily arXiv:1412.4561
  [astro-ph.HE]}}.

\bibitem{White19a}
C.~J. {White}, J.~M. {Stone}, and E.~{Quataert}, ``{A Resolution Study of
  Magnetically Arrested Disks},''
  \href{https://dx.doi.org/10.3847/1538-4357/ab0c0c}{{\em \apj} {\bfseries 874}
  no.~2, (Apr., 2019) 168}, \href{https://arxiv.org/abs/1903.01509}{{\ttfamily
  arXiv:1903.01509 [astro-ph.HE]}}.

\bibitem{Curd23}
B.~{Curd} and R.~{Narayan}, ``{GRRMHD simulations of MAD accretion discs
  declining from super-Eddington to sub-Eddington accretion rates},''
  \href{https://dx.doi.org/10.1093/mnras/stac3330}{{\em \mnras} {\bfseries 518}
  no.~3, (Jan., 2023) 3441--3461},
  \href{https://arxiv.org/abs/2209.12081}{{\ttfamily arXiv:2209.12081
  [astro-ph.HE]}}.

\bibitem{Scepi24}
N.~{Scepi}, M.~C. {Begelman}, and J.~{Dexter}, ``{Magnetic support, wind-driven
  accretion, coronal heating, and fast outflows in a thin magnetically arrested
  disc},'' \href{https://dx.doi.org/10.1093/mnras/stad3299}{{\em \mnras}
  {\bfseries 527} no.~1, (Jan., 2024) 1424--1443},
  \href{https://arxiv.org/abs/2302.10226}{{\ttfamily arXiv:2302.10226
  [astro-ph.HE]}}.

\bibitem{Fragile07}
P.~C. {Fragile}, O.~M. {Blaes}, P.~{Anninos}, and J.~D. {Salmonson}, ``{Global
  General Relativistic Magnetohydrodynamic Simulation of a Tilted Black Hole
  Accretion Disk},'' \href{https://dx.doi.org/10.1086/521092}{{\em \apj}
  {\bfseries 668} no.~1, (Oct., 2007) 417--429},
  \href{https://arxiv.org/abs/0706.4303}{{\ttfamily arXiv:0706.4303
  [astro-ph]}}.

\bibitem{McKinney13}
J.~C. {McKinney}, A.~{Tchekhovskoy}, and R.~D. {Blandford}, ``{Alignment of
  Magnetized Accretion Disks and Relativistic Jets with Spinning Black
  Holes},'' \href{https://dx.doi.org/10.1126/science.1230811}{{\em Science}
  {\bfseries 339} no.~6115, (Jan., 2013) 49},
  \href{https://arxiv.org/abs/1211.3651}{{\ttfamily arXiv:1211.3651
  [astro-ph.CO]}}.

\bibitem{Morales14}
D.~{Morales Teixeira}, P.~C. {Fragile}, V.~V. {Zhuravlev}, and P.~B. {Ivanov},
  ``{Conservative GRMHD Simulations of Moderately Thin, Tilted Accretion
  Disks},'' \href{https://dx.doi.org/10.1088/0004-637X/796/2/103}{{\em \apj}
  {\bfseries 796} no.~2, (Dec., 2014) 103},
  \href{https://arxiv.org/abs/1406.5514}{{\ttfamily arXiv:1406.5514
  [astro-ph.HE]}}.

\bibitem{Liska18}
M.~{Liska}, C.~{Hesp}, A.~{Tchekhovskoy}, A.~{Ingram}, M.~{van der Klis}, and
  S.~{Markoff}, ``{Formation of precessing jets by tilted black hole discs in
  3D general relativistic MHD simulations},''
  \href{https://dx.doi.org/10.1093/mnrasl/slx174}{{\em \mnras} {\bfseries 474}
  no.~1, (Feb., 2018) L81--L85},
  \href{https://arxiv.org/abs/1707.06619}{{\ttfamily arXiv:1707.06619
  [astro-ph.HE]}}.

\bibitem{Liska19}
M.~{Liska}, A.~{Tchekhovskoy}, A.~{Ingram}, and M.~{van der Klis},
  ``{Bardeen-Petterson alignment, jets, and magnetic truncation in GRMHD
  simulations of tilted thin accretion discs},''
  \href{https://dx.doi.org/10.1093/mnras/stz834}{{\em \mnras} {\bfseries 487}
  no.~1, (July, 2019) 550--561},
  \href{https://arxiv.org/abs/1810.00883}{{\ttfamily arXiv:1810.00883
  [astro-ph.HE]}}.

\bibitem{White19b}
C.~J. {White}, E.~{Quataert}, and O.~{Blaes}, ``{Tilted Disks around Black
  Holes: A Numerical Parameter Survey for Spin and Inclination Angle},''
  \href{https://dx.doi.org/10.3847/1538-4357/ab089e}{{\em \apj} {\bfseries 878}
  no.~1, (June, 2019) 51}, \href{https://arxiv.org/abs/1902.09662}{{\ttfamily
  arXiv:1902.09662 [astro-ph.HE]}}.

\bibitem{Takahashi16}
H.~R. {Takahashi}, K.~{Ohsuga}, T.~{Kawashima}, and Y.~{Sekiguchi},
  ``{Formation of Overheated Regions and Truncated Disks around Black Holes:
  Three-dimensional General Relativistic Radiation-magnetohydrodynamics
  Simulations},'' \href{https://dx.doi.org/10.3847/0004-637X/826/1/23}{{\em
  \apj} {\bfseries 826} no.~1, (July, 2016) 23},
  \href{https://arxiv.org/abs/1605.04992}{{\ttfamily arXiv:1605.04992
  [astro-ph.GA]}}.

\bibitem{Liska22a}
M.~T.~P. {Liska}, G.~{Musoke}, A.~{Tchekhovskoy}, O.~{Porth}, and A.~M.
  {Beloborodov}, ``{Formation of Magnetically Truncated Accretion Disks in 3D
  Radiation-transport Two-temperature GRMHD Simulations},''
  \href{https://dx.doi.org/10.3847/2041-8213/ac84db}{{\em \apjl} {\bfseries
  935} no.~1, (Aug., 2022) L1},
  \href{https://arxiv.org/abs/2201.03526}{{\ttfamily arXiv:2201.03526
  [astro-ph.HE]}}.

\bibitem{Bollimpalli24}
D.~A. {Bollimpalli}, P.~C. {Fragile}, J.~W. {Dewberry}, and W.~{Klu{\'z}niak},
  ``{Truncated, tilted discs as a possible source of Quasi-Periodic
  Oscillations},'' \href{https://dx.doi.org/10.1093/mnras/stad3975}{{\em
  \mnras} {\bfseries 528} no.~2, (Feb., 2024) 1142--1157},
  \href{https://arxiv.org/abs/2312.14876}{{\ttfamily arXiv:2312.14876
  [astro-ph.HE]}}.

\bibitem{Dionys2013}
K.~{Dionysopoulou}, D.~{Alic}, C.~{Palenzuela}, L.~{Rezzolla}, and
  B.~{Giacomazzo}, ``{General-relativistic resistive magnetohydrodynamics in
  three dimensions: Formulation and tests},''
  \href{https://dx.doi.org/10.1103/PhysRevD.88.044020}{{\em \prd} {\bfseries
  88} no.~4, (Aug., 2013) 044020},
  \href{https://arxiv.org/abs/1208.3487}{{\ttfamily arXiv:1208.3487 [gr-qc]}}.

\bibitem{Foucart2016}
F.~{Foucart}, M.~{Chandra}, C.~F. {Gammie}, and E.~{Quataert}, ``{Evolution of
  accretion discs around a kerr black hole using extended
  magnetohydrodynamics},'' \href{https://dx.doi.org/10.1093/mnras/stv2687}{{\em
  \mnras} {\bfseries 456} no.~2, (Feb., 2016) 1332--1345},
  \href{https://arxiv.org/abs/1511.04445}{{\ttfamily arXiv:1511.04445
  [astro-ph.HE]}}.

\bibitem{Mizuno2018}
Y.~{Mizuno}, Z.~{Younsi}, C.~M. {Fromm}, O.~{Porth}, M.~{De Laurentis},
  H.~{Olivares}, H.~{Falcke}, M.~{Kramer}, and L.~{Rezzolla}, ``{The current
  ability to test theories of gravity with black hole shadows},''
  \href{https://dx.doi.org/10.1038/s41550-018-0449-5}{{\em Nature Astronomy}
  {\bfseries 2} (Apr., 2018) 585--590},
  \href{https://arxiv.org/abs/1804.05812}{{\ttfamily arXiv:1804.05812
  [astro-ph.GA]}}.

\bibitem{Chatterjee2023}
K.~{Chatterjee}, P.~{Kocherlakota}, Z.~{Younsi}, and R.~{Narayan}, ``{Energy
  Extraction from Spinning Stringy Black Holes},''
  \href{https://dx.doi.org/10.48550/arXiv.2310.20040}{{\em arXiv e-prints}
  (Oct., 2023) arXiv:2310.20040},
  \href{https://arxiv.org/abs/2310.20040}{{\ttfamily arXiv:2310.20040
  [gr-qc]}}.

\bibitem{Abramowicz13}
M.~A. {Abramowicz} and P.~C. {Fragile}, ``{Foundations of Black Hole Accretion
  Disk Theory},'' \href{https://dx.doi.org/10.12942/lrr-2013-1}{{\em Living
  Reviews in Relativity} {\bfseries 16} no.~1, (Dec., 2013) 1},
  \href{https://arxiv.org/abs/1104.5499}{{\ttfamily arXiv:1104.5499
  [astro-ph.HE]}}.

\bibitem{McKinney2017}
J.~C. {McKinney}, J.~{Chluba}, M.~{Wielgus}, R.~{Narayan}, and A.~{Sadowski},
  ``{Double Compton and Cyclo-Synchrotron in Super-Eddington Discs, Magnetized
  Coronae, and Jets},'' \href{https://dx.doi.org/10.1093/mnras/stx227}{{\em
  \mnras} {\bfseries 467} no.~2, (May, 2017) 2241--2265},
  \href{https://arxiv.org/abs/1608.08627}{{\ttfamily arXiv:1608.08627
  [astro-ph.HE]}}.

\bibitem{Davis09}
S.~W. {Davis}, O.~M. {Blaes}, S.~{Hirose}, and J.~H. {Krolik}, ``{The Effects
  of Magnetic Fields and Inhomogeneities on Accretion Disk Spectra and
  Polarization},'' \href{https://dx.doi.org/10.1088/0004-637X/703/1/569}{{\em
  \apj} {\bfseries 703} no.~1, (Sept., 2009) 569--584},
  \href{https://arxiv.org/abs/0908.0505}{{\ttfamily arXiv:0908.0505
  [astro-ph.HE]}}.

\bibitem{Dolence09}
J.~C. {Dolence}, C.~F. {Gammie}, M.~{Mo{\'s}cibrodzka}, and P.~K. {Leung},
  ``{grmonty: A Monte Carlo Code for Relativistic Radiative Transport},''
  \href{https://dx.doi.org/10.1088/0067-0049/184/2/387}{{\em \apjs} {\bfseries
  184} no.~2, (Oct., 2009) 387--397},
  \href{https://arxiv.org/abs/0909.0708}{{\ttfamily arXiv:0909.0708
  [astro-ph.HE]}}.

\bibitem{Schnittman13b}
J.~D. {Schnittman} and J.~H. {Krolik}, ``{A Monte Carlo Code for Relativistic
  Radiation Transport around Kerr Black Holes},''
  \href{https://dx.doi.org/10.1088/0004-637X/777/1/11}{{\em \apj} {\bfseries
  777} no.~1, (Nov., 2013) 11},
  \href{https://arxiv.org/abs/1302.3214}{{\ttfamily arXiv:1302.3214
  [astro-ph.HE]}}.

\bibitem{Dexter16}
J.~{Dexter}, ``{A public code for general relativistic, polarised radiative
  transfer around spinning black holes},''
  \href{https://dx.doi.org/10.1093/mnras/stw1526}{{\em \mnras} {\bfseries 462}
  no.~1, (Oct., 2016) 115--136},
  \href{https://arxiv.org/abs/1602.03184}{{\ttfamily arXiv:1602.03184
  [astro-ph.HE]}}.

\bibitem{Narayan16}
R.~{Narayan}, Y.~{Zhu}, D.~{Psaltis}, and A.~{Sadowski}, ``{HEROIC: 3D general
  relativistic radiative post-processor with comptonization for black hole
  accretion discs},'' \href{https://dx.doi.org/10.1093/mnras/stv2979}{{\em
  \mnras} {\bfseries 457} no.~1, (Mar., 2016) 608--628},
  \href{https://arxiv.org/abs/1510.04208}{{\ttfamily arXiv:1510.04208
  [astro-ph.HE]}}.

\bibitem{Bronzwaer18}
T.~{Bronzwaer}, J.~{Davelaar}, Z.~{Younsi}, M.~{Mo{\'s}cibrodzka}, H.~{Falcke},
  M.~{Kramer}, and L.~{Rezzolla}, ``{RAPTOR. I. Time-dependent radiative
  transfer in arbitrary spacetimes},''
  \href{https://dx.doi.org/10.1051/0004-6361/201732149}{{\em \aap} {\bfseries
  613} (May, 2018) A2}, \href{https://arxiv.org/abs/1801.10452}{{\ttfamily
  arXiv:1801.10452 [astro-ph.HE]}}.

\bibitem{Roth25}
N.~{Roth}, P.~{Anninos}, P.~C. {Fragile}, and D.~{Pickrel}, ``{X-Ray Spectra
  from General Relativistic Radiation Magnetohydrodynamic Simulations of Thin
  Disks},'' \href{https://dx.doi.org/10.3847/1538-4357/adb1c1}{{\em \apj}
  {\bfseries 981} no.~2, (Mar., 2025) 144},
  \href{https://arxiv.org/abs/2501.18040}{{\ttfamily arXiv:2501.18040
  [astro-ph.HE]}}.

\bibitem{Schnittman13a}
J.~D. {Schnittman}, J.~H. {Krolik}, and S.~C. {Noble}, ``{X-Ray Spectra from
  Magnetohydrodynamic Simulations of Accreting Black Holes},''
  \href{https://dx.doi.org/10.1088/0004-637X/769/2/156}{{\em \apj} {\bfseries
  769} no.~2, (June, 2013) 156},
  \href{https://arxiv.org/abs/1207.2693}{{\ttfamily arXiv:1207.2693
  [astro-ph.HE]}}.

\bibitem{Kinch19}
B.~E. {Kinch}, J.~D. {Schnittman}, T.~R. {Kallman}, and J.~H. {Krolik},
  ``{Predicting the X-Ray Spectra of Stellar-mass Black Holes from
  Simulations},'' \href{https://dx.doi.org/10.3847/1538-4357/ab05d5}{{\em \apj}
  {\bfseries 873} no.~1, (Mar., 2019) 71}.

\bibitem{Wielgus2022}
M.~{Wielgus}, D.~{Lan{\v{c}}ov{\'a}}, O.~{Straub}, W.~{Klu{\'z}niak},
  R.~{Narayan}, D.~{Abarca}, A.~{R{\'o}{\.z}a{\'n}ska}, F.~{Vincent},
  G.~{T{\"o}r{\"o}k}, and M.~{Abramowicz}, ``{Observational properties of puffy
  discs: radiative GRMHD spectra of mildly sub-Eddington accretion},''
  \href{https://dx.doi.org/10.1093/mnras/stac1317}{{\em \mnras} {\bfseries 514}
  no.~1, (July, 2022) 780--789},
  \href{https://arxiv.org/abs/2202.08831}{{\ttfamily arXiv:2202.08831
  [astro-ph.HE]}}.

\bibitem{Mills24}
B.~S. {Mills}, S.~W. {Davis}, Y.-F. {Jiang}, and M.~J. {Middleton}, ``{Spectral
  Calculations of 3D Radiation Magnetohydrodynamic Simulations of
  Super-Eddington Accretion onto a Stellar-mass Black Hole},''
  \href{https://dx.doi.org/10.3847/1538-4357/ad6b21}{{\em \apj} {\bfseries 974}
  no.~2, (Oct., 2024) 166}, \href{https://arxiv.org/abs/2304.07977}{{\ttfamily
  arXiv:2304.07977 [astro-ph.HE]}}.

\bibitem{Moscibrodzka2024}
M.~{Moscibrodzka}, ``{What is the hard spectral state in X-ray binaries?
  Insights from GRRMHD accretion flows simulations and polarization of their
  X-ray emission},'' \href{https://dx.doi.org/10.1007/s10509-024-04333-3}{{\em
  \apss} {\bfseries 369} no.~7, (July, 2024) 68},
  \href{https://arxiv.org/abs/2309.09087}{{\ttfamily arXiv:2309.09087
  [astro-ph.HE]}}.

\bibitem{Shcherbakov12}
R.~V. {Shcherbakov}, R.~F. {Penna}, and J.~C. {McKinney}, ``{Sagittarius A*
  Accretion Flow and Black Hole Parameters from General Relativistic Dynamical
  and Polarized Radiative Modeling},''
  \href{https://dx.doi.org/10.1088/0004-637X/755/2/133}{{\em \apj} {\bfseries
  755} no.~2, (Aug., 2012) 133},
  \href{https://arxiv.org/abs/1007.4832}{{\ttfamily arXiv:1007.4832
  [astro-ph.HE]}}.

\bibitem{Chan15}
C.-K. {Chan}, D.~{Psaltis}, F.~{{\"O}zel}, R.~{Narayan}, and A.~{Sadowski},
  ``{The Power of Imaging: Constraining the Plasma Properties of GRMHD
  Simulations using EHT Observations of Sgr A*},''
  \href{https://dx.doi.org/10.1088/0004-637X/799/1/1}{{\em \apj} {\bfseries
  799} no.~1, (Jan., 2015) 1},
  \href{https://arxiv.org/abs/1410.3492}{{\ttfamily arXiv:1410.3492
  [astro-ph.HE]}}.

\bibitem{EHTC_M87_p5}
{Event Horizon Telescope Collaboration}, ``{First M87 Event Horizon Telescope
  Results. V. Physical Origin of the Asymmetric Ring},''
  \href{https://dx.doi.org/10.3847/2041-8213/ab0f43}{{\em \apjl} {\bfseries
  875} no.~1, (Apr., 2019) L5},
  \href{https://arxiv.org/abs/1906.11242}{{\ttfamily arXiv:1906.11242
  [astro-ph.GA]}}.

\bibitem{EHTC_SgrA_p5}
{Event Horizon Telescope Collaboration}, ``{First Sagittarius A* Event Horizon
  Telescope Results. V. Testing Astrophysical Models of the Galactic Center
  Black Hole},'' \href{https://dx.doi.org/10.3847/2041-8213/ac6672}{{\em \apjl}
  {\bfseries 930} no.~2, (May, 2022) L16}.

\bibitem{Kinch16}
B.~E. {Kinch}, J.~D. {Schnittman}, T.~R. {Kallman}, and J.~H. {Krolik}, ``{Fe
  K{\ensuremath{\alpha}} Profiles from Simulations of Accreting Black Holes},''
  \href{https://dx.doi.org/10.3847/0004-637X/826/1/52}{{\em \apj} {\bfseries
  826} no.~1, (July, 2016) 52},
  \href{https://arxiv.org/abs/1604.01126}{{\ttfamily arXiv:1604.01126
  [astro-ph.HE]}}.

\bibitem{Noble09}
S.~C. {Noble}, J.~H. {Krolik}, and J.~F. {Hawley}, ``{Direct Calculation of the
  Radiative Efficiency of an Accretion Disk Around a Black Hole},''
  \href{https://dx.doi.org/10.1088/0004-637X/692/1/411}{{\em \apj} {\bfseries
  692} no.~1, (Feb., 2009) 411--421},
  \href{https://arxiv.org/abs/0808.3140}{{\ttfamily arXiv:0808.3140
  [astro-ph]}}.

\bibitem{Kinch21}
B.~E. {Kinch}, J.~D. {Schnittman}, S.~C. {Noble}, T.~R. {Kallman}, and J.~H.
  {Krolik}, ``{Spin and Accretion Rate Dependence of Black Hole X-Ray
  Spectra},'' \href{https://dx.doi.org/10.3847/1538-4357/ac2b9a}{{\em \apj}
  {\bfseries 922} no.~2, (Dec., 2021) 270},
  \href{https://arxiv.org/abs/2105.09408}{{\ttfamily arXiv:2105.09408
  [astro-ph.HE]}}.

\bibitem{Schnittman06a}
J.~D. {Schnittman} and L.~{Rezzolla}, ``{Quasi-periodic Oscillations in the
  X-Ray Light Curves from Relativistic Tori},''
  \href{https://dx.doi.org/10.1086/500545}{{\em \apjl} {\bfseries 637} no.~2,
  (Feb., 2006) L113--L116},
  \href{https://arxiv.org/abs/astro-ph/0506702}{{\ttfamily
  arXiv:astro-ph/0506702 [astro-ph]}}.

\bibitem{Schnittman06b}
J.~D. {Schnittman}, J.~H. {Krolik}, and J.~F. {Hawley}, ``{Light Curves from an
  MHD Simulation of a Black Hole Accretion Disk},''
  \href{https://dx.doi.org/10.1086/507421}{{\em \apj} {\bfseries 651} no.~2,
  (Nov., 2006) 1031--1048},
  \href{https://arxiv.org/abs/astro-ph/0606615}{{\ttfamily
  arXiv:astro-ph/0606615 [astro-ph]}}.

\bibitem{Mishra17}
B.~{Mishra}, F.~H. {Vincent}, A.~{Manousakis}, P.~C. {Fragile}, T.~{Paumard},
  and W.~{Klu{\'z}niak}, ``{Quasi-periodic oscillations from relativistic
  ray-traced hydrodynamical tori},''
  \href{https://dx.doi.org/10.1093/mnras/stx299}{{\em \mnras} {\bfseries 467}
  no.~4, (June, 2017) 4036--4049},
  \href{https://arxiv.org/abs/1510.07414}{{\ttfamily arXiv:1510.07414
  [astro-ph.HE]}}.

\bibitem{Bollimpalli20}
D.~A. {Bollimpalli}, R.~{Mahmoud}, C.~{Done}, P.~C. {Fragile},
  W.~{Klu{\'z}niak}, R.~{Narayan}, and C.~J. {White}, ``{Looking for the
  underlying cause of black hole X-ray variability in GRMHD simulations},''
  \href{https://dx.doi.org/10.1093/mnras/staa1808}{{\em \mnras} {\bfseries 496}
  no.~3, (Aug., 2020) 3808--3828},
  \href{https://arxiv.org/abs/2006.00755}{{\ttfamily arXiv:2006.00755
  [astro-ph.HE]}}.

\bibitem{Salas2025}
L.~D.~S. {Salas}, M.~T.~P. {Liska}, S.~B. {Markoff}, K.~{Chatterjee},
  G.~{Musoke}, O.~{Porth}, B.~{Ripperda}, D.~{Yoon}, and W.~{Mulaudzi},
  ``{Two-temperature treatments in magnetically arrested disc GRMHD simulations
  more accurately predict light curves of Sagittarius A*},''
  \href{https://dx.doi.org/10.1093/mnras/staf240}{{\em \mnras} {\bfseries 538}
  no.~2, (Apr., 2025) 698--710},
  \href{https://arxiv.org/abs/2411.09556}{{\ttfamily arXiv:2411.09556
  [astro-ph.HE]}}.

\bibitem{Baade1956}
W.~{Baade}, ``{Polarization in the Jet of Messier 87.},''
  \href{https://dx.doi.org/10.1086/146194}{{\em \apj} {\bfseries 123} (May,
  1956) 550--551}.

\bibitem{Angel1980}
J.~R.~P. {Angel} and H.~S. {Stockman}, ``{Optical and infrared polarization of
  active extragalactic objects},''
  \href{https://dx.doi.org/10.1146/annurev.aa.18.090180.001541}{{\em \araa}
  {\bfseries 18} (Jan., 1980) 321--361}.

\bibitem{Wardle1974}
J.~F.~C. {Wardle} and P.~P. {Kronberg}, ``{The linear polarization of
  quasi-stellar radio sources at 3.71 and 11.1 centimeters.},''
  \href{https://dx.doi.org/10.1086/153240}{{\em \apj} {\bfseries 194} (Jan.,
  1974) 249}.

\bibitem{Doroshenko22}
V.~{Doroshenko}, J.~{Poutanen}, {\em et~al.}, ``{Determination of X-ray pulsar
  geometry with IXPE polarimetry},''
  \href{https://dx.doi.org/10.1038/s41550-022-01799-5}{{\em Nature Astronomy}
  {\bfseries 6} (Dec., 2022) 1433--1443},
  \href{https://arxiv.org/abs/2206.07138}{{\ttfamily arXiv:2206.07138
  [astro-ph.HE]}}.

\bibitem{Krawczynski22}
H.~{Krawczynski}, F.~{Muleri}, {\em et~al.}, ``{Polarized x-rays constrain the
  disk-jet geometry in the black hole x-ray binary Cygnus X-1},''
  \href{https://dx.doi.org/10.1126/science.add5399}{{\em Science} {\bfseries
  378} no.~6620, (Nov., 2022) 650--654},
  \href{https://arxiv.org/abs/2206.09972}{{\ttfamily arXiv:2206.09972
  [astro-ph.HE]}}.

\bibitem{Podgorny23}
J.~{Podgorn{\'y}}, L.~{Marra}, {\em et~al.}, ``{The first X-ray polarimetric
  observation of the black hole binary LMC X-1},''
  \href{https://dx.doi.org/10.1093/mnras/stad3103}{{\em \mnras} {\bfseries 526}
  no.~4, (Dec., 2023) 5964--5975},
  \href{https://arxiv.org/abs/2303.12034}{{\ttfamily arXiv:2303.12034
  [astro-ph.HE]}}.

\bibitem{Weisskopf22}
M.~C. {Weisskopf}, P.~{Soffitta}, {\em et~al.}, ``{The Imaging X-Ray
  Polarimetry Explorer (IXPE): Pre-Launch},''
  \href{https://dx.doi.org/10.1117/1.JATIS.8.2.026002}{{\em Journal of
  Astronomical Telescopes, Instruments, and Systems} {\bfseries 8} no.~2,
  (Apr., 2022) 026002}, \href{https://arxiv.org/abs/2112.01269}{{\ttfamily
  arXiv:2112.01269 [astro-ph.IM]}}.

\bibitem{Radhakrishna25}
R.~{V}, A.~{Tyagi}, {\em et~al.}, ``{X-Ray spectroscopy and timing (XSPECT)
  experiment on XPoSat -- instrument configuration and science prospects},''
  \href{https://dx.doi.org/10.48550/arXiv.2505.20061}{{\em arXiv e-prints}
  (May, 2025) arXiv:2505.20061},
  \href{https://arxiv.org/abs/2505.20061}{{\ttfamily arXiv:2505.20061
  [astro-ph.IM]}}.

\bibitem{Zhang19}
S.~{Zhang}, A.~{Santangelo}, {\em et~al.}, ``{The enhanced X-ray Timing and
  Polarimetry mission{\textemdash}eXTP},''
  \href{https://dx.doi.org/10.1007/s11433-018-9309-2}{{\em Science China
  Physics, Mechanics, and Astronomy} {\bfseries 62} no.~2, (Feb., 2019) 29502},
  \href{https://arxiv.org/abs/1812.04020}{{\ttfamily arXiv:1812.04020
  [astro-ph.IM]}}.

\bibitem{Schnittman09}
J.~D. {Schnittman} and J.~H. {Krolik}, ``{X-ray Polarization from Accreting
  Black Holes: The Thermal State},''
  \href{https://dx.doi.org/10.1088/0004-637X/701/2/1175}{{\em \apj} {\bfseries
  701} no.~2, (Aug., 2009) 1175--1187},
  \href{https://arxiv.org/abs/0902.3982}{{\ttfamily arXiv:0902.3982
  [astro-ph.HE]}}.

\bibitem{Cheng16}
Y.~{Cheng}, D.~{Liu}, S.~{Nampalliwar}, and C.~{Bambi}, ``{X-ray
  spectropolarimetric signature of a warped disk around a stellar-mass black
  hole},'' \href{https://dx.doi.org/10.1088/0264-9381/33/12/125015}{{\em
  Classical and Quantum Gravity} {\bfseries 33} no.~12, (June, 2016) 125015},
  \href{https://arxiv.org/abs/1505.01562}{{\ttfamily arXiv:1505.01562
  [gr-qc]}}.

\bibitem{Farinelli23}
R.~{Farinelli}, S.~{Fabiani}, {\em et~al.}, ``{Accretion geometry of the
  neutron star low mass X-ray binary Cyg X-2 from X-ray polarization
  measurements},'' \href{https://dx.doi.org/10.1093/mnras/stac3726}{{\em
  \mnras} {\bfseries 519} no.~3, (Mar., 2023) 3681--3690},
  \href{https://arxiv.org/abs/2212.13119}{{\ttfamily arXiv:2212.13119
  [astro-ph.HE]}}.

\bibitem{Schnittman10}
J.~D. {Schnittman} and J.~H. {Krolik}, ``{X-ray Polarization from Accreting
  Black Holes: Coronal Emission},''
  \href{https://dx.doi.org/10.1088/0004-637X/712/2/908}{{\em \apj} {\bfseries
  712} no.~2, (Apr., 2010) 908--924},
  \href{https://arxiv.org/abs/0912.0907}{{\ttfamily arXiv:0912.0907
  [astro-ph.HE]}}.

\bibitem{Marinucci22}
A.~{Marinucci}, F.~{Muleri}, {\em et~al.}, ``{Polarization constraints on the
  X-ray corona in Seyfert Galaxies: MCG-05-23-16},''
  \href{https://dx.doi.org/10.1093/mnras/stac2634}{{\em \mnras} {\bfseries 516}
  no.~4, (Nov., 2022) 5907--5913},
  \href{https://arxiv.org/abs/2207.09338}{{\ttfamily arXiv:2207.09338
  [astro-ph.HE]}}.

\bibitem{Gianolli23}
V.~E. {Gianolli}, D.~E. {Kim}, {\em et~al.}, ``{Uncovering the geometry of the
  hot X-ray corona in the Seyfert galaxy NGC 4151 with IXPE},''
  \href{https://dx.doi.org/10.1093/mnras/stad1697}{{\em \mnras} {\bfseries 523}
  no.~3, (Aug., 2023) 4468--4476},
  \href{https://arxiv.org/abs/2303.12541}{{\ttfamily arXiv:2303.12541
  [astro-ph.GA]}}.

\bibitem{Marra2024}
L.~{Marra}, M.~{Brigitte}, {\em et~al.}, ``{IXPE observation confirms a high
  spin in the accreting black hole 4U 1957+115},''
  \href{https://dx.doi.org/10.1051/0004-6361/202348277}{{\em \aap} {\bfseries
  684} (Apr., 2024) A95}, \href{https://arxiv.org/abs/2310.11125}{{\ttfamily
  arXiv:2310.11125 [astro-ph.HE]}}.

\bibitem{Saurabh2025}
{Saurabh}, M.~{Wielgus}, A.~{Tursunov}, A.~P. {Lobanov}, and R.~{Emami},
  ``{Semi-analytic studies of the accretion disk and magnetic field geometry in
  M 87*},'' \href{https://dx.doi.org/10.1051/0004-6361/202556880}{{\em \aap}
  {\bfseries 705} (Jan., 2026) A166},
  \href{https://arxiv.org/abs/2508.11760}{{\ttfamily arXiv:2508.11760
  [astro-ph.HE]}}.

\bibitem{Toth00}
G.~{T{\'o}th}, ``{The {\ensuremath{\nabla}}{\textperiodcentered} B=0 Constraint
  in Shock-Capturing Magnetohydrodynamics Codes},''
  \href{https://dx.doi.org/10.1006/jcph.2000.6519}{{\em Journal of
  Computational Physics} {\bfseries 161} no.~2, (July, 2000) 605--652}.

\bibitem{Thorne81}
K.~S. {Thorne}, ``{Relativistic radiative transfer - Moment formalisms},''
  \href{https://dx.doi.org/10.1093/mnras/194.2.439}{{\em \mnras} {\bfseries
  194} (Feb., 1981) 439--473}.

\bibitem{Shibata11}
M.~{Shibata}, K.~{Kiuchi}, Y.~{Sekiguchi}, and Y.~{Suwa}, ``{Truncated Moment
  Formalism for Radiation Hydrodynamics in Numerical Relativity},''
  \href{https://dx.doi.org/10.1143/PTP.125.1255}{{\em Progress of Theoretical
  Physics} {\bfseries 125} no.~6, (June, 2011) 1255--1287},
  \href{https://arxiv.org/abs/1104.3937}{{\ttfamily arXiv:1104.3937
  [astro-ph.HE]}}.

\bibitem{Anninos20}
P.~{Anninos} and P.~C. {Fragile}, ``{Multi-frequency General Relativistic
  Radiation-hydrodynamics with M$_{1}$ Closure},''
  \href{https://dx.doi.org/10.3847/1538-4357/abab9c}{{\em \apj} {\bfseries 900}
  no.~1, (Sept., 2020) 71}, \href{https://arxiv.org/abs/2007.12195}{{\ttfamily
  arXiv:2007.12195 [astro-ph.IM]}}.

\bibitem{Levermore84}
C.~D. {Levermore}, ``{Relating Eddington factors to flux limiters.},''
  \href{https://dx.doi.org/10.1016/0022-4073(84)90112-2}{{\em \jqsrt}
  {\bfseries 31} no.~2, (Feb., 1984) 149--160}.

\bibitem{Sadowski13b}
A.~{S{\k{a}}dowski}, R.~{Narayan}, A.~{Tchekhovskoy}, and Y.~{Zhu},
  ``{Semi-implicit scheme for treating radiation under M1 closure in general
  relativistic conservative fluid dynamics codes},''
  \href{https://dx.doi.org/10.1093/mnras/sts632}{{\em \mnras} {\bfseries 429}
  no.~4, (Mar., 2013) 3533--3550},
  \href{https://arxiv.org/abs/1212.5050}{{\ttfamily arXiv:1212.5050
  [astro-ph.HE]}}.

\bibitem{Fragile14}
P.~C. {Fragile}, A.~{Olejar}, and P.~{Anninos}, ``{Numerical Simulations of
  Optically Thick Accretion onto a Black Hole. II. Rotating Flow},''
  \href{https://dx.doi.org/10.1088/0004-637X/796/1/22}{{\em \apj} {\bfseries
  796} no.~1, (Nov., 2014) 22},
  \href{https://arxiv.org/abs/1408.4460}{{\ttfamily arXiv:1408.4460
  [astro-ph.IM]}}.

\bibitem{Davis20}
S.~W. {Davis} and C.~F. {Gammie}, ``{Covariant Radiative Transfer for Black
  Hole Spacetimes},'' \href{https://dx.doi.org/10.3847/1538-4357/ab5950}{{\em
  \apj} {\bfseries 888} no.~2, (Jan., 2020) 94},
  \href{https://arxiv.org/abs/1911.07950}{{\ttfamily arXiv:1911.07950
  [astro-ph.HE]}}.

\bibitem{Fragile23}
P.~C. {Fragile}, P.~{Anninos}, N.~{Roth}, and B.~{Mishra}, ``{Multifrequency
  General Relativistic Radiation Magnetohydrodynamic Simulations of Thin
  Disks},'' \href{https://dx.doi.org/10.3847/1538-4357/ad096b}{{\em \apj}
  {\bfseries 959} no.~1, (Dec., 2023) 59},
  \href{https://arxiv.org/abs/2311.00028}{{\ttfamily arXiv:2311.00028
  [astro-ph.HE]}}.

\bibitem{Anninos05}
P.~{Anninos}, P.~C. {Fragile}, and J.~D. {Salmonson}, ``{Cosmos++: Relativistic
  Magnetohydrodynamics on Unstructured Grids with Local Adaptive Refinement},''
  \href{https://dx.doi.org/10.1086/497294}{{\em \apj} {\bfseries 635} no.~1,
  (Dec., 2005) 723--740},
  \href{https://arxiv.org/abs/astro-ph/0509254}{{\ttfamily
  arXiv:astro-ph/0509254 [astro-ph]}}.

\bibitem{DelZanna07}
L.~{Del Zanna}, O.~{Zanotti}, N.~{Bucciantini}, and P.~{Londrillo}, ``{ECHO: a
  Eulerian conservative high-order scheme for general relativistic
  magnetohydrodynamics and magnetodynamics},''
  \href{https://dx.doi.org/10.1051/0004-6361:20077093}{{\em \aap} {\bfseries
  473} no.~1, (Oct., 2007) 11--30},
  \href{https://arxiv.org/abs/0704.3206}{{\ttfamily arXiv:0704.3206
  [astro-ph]}}.

\bibitem{White16}
C.~J. {White}, J.~M. {Stone}, and C.~F. {Gammie}, ``{An Extension of the
  Athena++ Code Framework for GRMHD Based on Advanced Riemann Solvers and
  Staggered-mesh Constrained Transport},''
  \href{https://dx.doi.org/10.3847/0067-0049/225/2/22}{{\em \apjs} {\bfseries
  225} no.~2, (Aug., 2016) 22},
  \href{https://arxiv.org/abs/1511.00943}{{\ttfamily arXiv:1511.00943
  [astro-ph.HE]}}.

\bibitem{Porth17}
O.~{Porth}, H.~{Olivares}, Y.~{Mizuno}, Z.~{Younsi}, L.~{Rezzolla},
  M.~{Moscibrodzka}, H.~{Falcke}, and M.~{Kramer}, ``{The black hole accretion
  code},'' \href{https://dx.doi.org/10.1186/s40668-017-0020-2}{{\em
  Computational Astrophysics and Cosmology} {\bfseries 4} no.~1, (May, 2017)
  1}, \href{https://arxiv.org/abs/1611.09720}{{\ttfamily arXiv:1611.09720
  [gr-qc]}}.

\bibitem{Prather21}
B.~{Prather}, G.~{Wong}, V.~{Dhruv}, B.~{Ryan}, J.~{Dolence}, S.~{Ressler}, and
  C.~{Gammie}, ``{iharm3D: Vectorized General Relativistic
  Magnetohydrodynamics},'' \href{https://dx.doi.org/10.21105/joss.03336}{{\em
  The Journal of Open Source Software} {\bfseries 6} no.~66, (Oct., 2021)
  3336}, \href{https://arxiv.org/abs/2110.10191}{{\ttfamily arXiv:2110.10191
  [astro-ph.HE]}}.

\bibitem{Liska22b}
M.~T.~P. {Liska}, K.~{Chatterjee}, {\em et~al.}, ``{H-AMR: A New
  GPU-accelerated GRMHD Code for Exascale Computing with 3D Adaptive Mesh
  Refinement and Local Adaptive Time Stepping},''
  \href{https://dx.doi.org/10.3847/1538-4365/ac9966}{{\em \apjs} {\bfseries
  263} no.~2, (Dec., 2022) 26},
  \href{https://arxiv.org/abs/1912.10192}{{\ttfamily arXiv:1912.10192
  [astro-ph.HE]}}.

\bibitem{Porth19}
O.~{Porth}, K.~{Chatterjee}, {\em et~al.}, ``{The Event Horizon General
  Relativistic Magnetohydrodynamic Code Comparison Project},''
  \href{https://dx.doi.org/10.3847/1538-4365/ab29fd}{{\em \apjs} {\bfseries
  243} no.~2, (Aug., 2019) 26},
  \href{https://arxiv.org/abs/1904.04923}{{\ttfamily arXiv:1904.04923
  [astro-ph.HE]}}.

\bibitem{Dexter2009}
J.~Dexter and E.~Agol, ``A {{FAST NEW PUBLIC CODE FOR COMPUTING PHOTON ORBITS
  IN A KERR SPACETIME}},''.

\bibitem{Gralla2020}
S.~E. Gralla and A.~Lupsasca, ``Null geodesics of the {Kerr} exterior,''
  \href{https://dx.doi.org/10.1103/PhysRevD.101.044032}{{\em Physical Review D}
  {\bfseries 101} no.~4, (Feb., 2020) 044032}.
  \url{https://link.aps.org/doi/10.1103/PhysRevD.101.044032}.

\bibitem{Broderick2004}
A.~Broderick and R.~Blandford, ``Covariant magnetoionic theory - {{II}}.
  {{Radiative}} transfer,''.

\bibitem{Gammie2012}
C.~F. Gammie and P.~K. Leung, ``A {{Formalism}} for {{Covariant Polarized
  Radiative Transport}} by {{Ray Tracing}},''.

\bibitem{Luminet79}
J.~P. {Luminet}, ``{Image of a spherical black hole with thin accretion
  disk.},'' {\em \aap} {\bfseries 75} (May, 1979) 228--235.

\bibitem{EHTC_M87_p1}
{Event Horizon Telescope Collaboration}, ``{First M87 Event Horizon Telescope
  Results. I. The Shadow of the Supermassive Black Hole},''
  \href{https://dx.doi.org/10.3847/2041-8213/ab0ec7}{{\em \apjl} {\bfseries
  875} no.~1, (Apr., 2019) L1},
  \href{https://arxiv.org/abs/1906.11238}{{\ttfamily arXiv:1906.11238
  [astro-ph.GA]}}.

\bibitem{EHTC_SgrA_p1}
{Event Horizon Telescope Collaboration}, ``{First Sagittarius A* Event Horizon
  Telescope Results. I. The Shadow of the Supermassive Black Hole in the Center
  of the Milky Way},'' \href{https://dx.doi.org/10.3847/2041-8213/ac6674}{{\em
  \apjl} {\bfseries 930} no.~2, (May, 2022) L12}.

\bibitem{White2023}
C.~J. White, P.~D. Mullen, Y.-F. Jiang, S.~W. Davis, J.~M. Stone, V.~Morozova,
  and L.~Zhang, ``An {{Extension}} of the {{Athena}}++ {{Code Framework}} for
  {{Radiation-magnetohydrodynamics}} in {{General Relativity Using}} a
  {{Finite-solid-angle Discretization}},''.

\bibitem{Yarza2020}
R.~Yarza, G.~N. Wong, B.~R. Ryan, and C.~F. Gammie, ``Bremsstrahlung in
  {{GRMHD}} models of accreting black holes,''
  \href{https://arxiv.org/abs/2006.01145}{{\ttfamily 2006.01145}}.

\bibitem{Kawashima2023}
T.~Kawashima, K.~Ohsuga, and H.~R. Takahashi, ``{{RAIKOU}}: {{A
  General Relativistic}}, {{Multiwavelength Radiative Transfer Code}},''.

\bibitem{Pelle2022}
J.~Pelle, O.~Reula, F.~Carrasco, and C.~Bederian, ``Skylight: A new code for
  general-relativistic ray-tracing and radiative transfer in arbitrary
  space--times,''.

\bibitem{Moscibrodzka2020}
M.~Mo\'scibrodzka, ``General relativistic polarized radiative transfer with
  inverse-{{Compton}} scatterings,''.

\bibitem{Stark77}
R.~F. {Stark} and P.~A. {Connors}, ``{Observational test for the existence of a
  rotating black hole in Cyg X-1},''
  \href{https://dx.doi.org/10.1038/266429a0}{{\em \nat} {\bfseries 266}
  no.~5601, (Mar., 1977) 429--430}.

\bibitem{Connors80}
P.~A. {Connors}, T.~{Piran}, and R.~F. {Stark}, ``{Polarization features of
  X-ray radiation emitted near black holes.},''
  \href{https://dx.doi.org/10.1086/157627}{{\em \apj} {\bfseries 235} (Jan.,
  1980) 224--244}.

\bibitem{Laor90}
A.~{Laor}, H.~{Netzer}, and T.~{Piran}, ``{Massive thin accretion discs. II -
  Polarization},'' \href{https://dx.doi.org/10.1093/mnras/242.4.560}{{\em
  \mnras} {\bfseries 242} (Feb., 1990) 560--569}.

\bibitem{Dovciak08}
M.~{Dov{\v{c}}iak}, F.~{Muleri}, R.~W. {Goosmann}, V.~{Karas}, and G.~{Matt},
  ``{Thermal disc emission from a rotating black hole: X-ray polarization
  signatures},''
  \href{https://dx.doi.org/10.1111/j.1365-2966.2008.13872.x}{{\em \mnras}
  {\bfseries 391} no.~1, (Nov., 2008) 32--38},
  \href{https://arxiv.org/abs/0809.0418}{{\ttfamily arXiv:0809.0418
  [astro-ph]}}.

\bibitem{Novikov73}
I.~D. {Novikov} and K.~S. {Thorne}, ``{Astrophysics of black holes.},'' in {\em
  Black Holes (Les Astres Occlus)}, C.~{Dewitt} and B.~S. {Dewitt}, eds.,
  pp.~343--450.
\newblock Jan., 1973.

\bibitem{Chandrasekhar60}
S.~{Chandrasekhar}, {\em {Radiative transfer}}.
\newblock 1960.

\bibitem{Cunningham76}
C.~{Cunningham}, ``{Returning radiation in accretion disks around black
  holes.},'' \href{https://dx.doi.org/10.1086/154636}{{\em \apj} {\bfseries
  208} (Sept., 1976) 534--549}.

\bibitem{Kinch20}
B.~E. {Kinch}, S.~C. {Noble}, J.~D. {Schnittman}, and J.~H. {Krolik},
  ``{Inverse Compton Cooling in the Coronae of Simulated Black Hole Accretion
  Flows},'' \href{https://dx.doi.org/10.3847/1538-4357/abc176}{{\em \apj}
  {\bfseries 904} no.~2, (Dec., 2020) 117},
  \href{https://arxiv.org/abs/2009.01914}{{\ttfamily arXiv:2009.01914
  [astro-ph.HE]}}.

\bibitem{Tamborra18}
F.~{Tamborra}, G.~{Matt}, S.~{Bianchi}, and M.~{Dov{\v{c}}iak}, ``{MoCA: A
  Monte Carlo code for Comptonisation in Astrophysics. I. Description of the
  code and first results},''
  \href{https://dx.doi.org/10.1051/0004-6361/201732023}{{\em \aap} {\bfseries
  619} (Nov., 2018) A105}, \href{https://arxiv.org/abs/1808.07399}{{\ttfamily
  arXiv:1808.07399 [astro-ph.HE]}}.

\bibitem{Broderick10}
A.~E. {Broderick} and J.~C. {McKinney}, ``{Parsec-scale Faraday Rotation
  Measures from General Relativistic Magnetohydrodynamic Simulations of Active
  Galactic Nucleus Jets},''
  \href{https://dx.doi.org/10.1088/0004-637X/725/1/750}{{\em \apj} {\bfseries
  725} no.~1, (Dec., 2010) 750--773},
  \href{https://arxiv.org/abs/1006.5015}{{\ttfamily arXiv:1006.5015
  [astro-ph.HE]}}.

\bibitem{Gold17}
R.~{Gold}, J.~C. {McKinney}, M.~D. {Johnson}, and S.~S. {Doeleman}, ``{Probing
  the Magnetic Field Structure in Sgr A* on Black Hole Horizon Scales with
  Polarized Radiative Transfer Simulations},''
  \href{https://dx.doi.org/10.3847/1538-4357/aa6193}{{\em \apj} {\bfseries 837}
  no.~2, (Mar., 2017) 180}, \href{https://arxiv.org/abs/1601.05550}{{\ttfamily
  arXiv:1601.05550 [astro-ph.HE]}}.

\bibitem{Moscibrodzka17}
M.~{Mo{\'s}cibrodzka}, J.~{Dexter}, J.~{Davelaar}, and H.~{Falcke}, ``{Faraday
  rotation in GRMHD simulations of the jet launching zone of M87},''
  \href{https://dx.doi.org/10.1093/mnras/stx587}{{\em \mnras} {\bfseries 468}
  no.~2, (June, 2017) 2214--2221},
  \href{https://arxiv.org/abs/1703.02390}{{\ttfamily arXiv:1703.02390
  [astro-ph.HE]}}.

\bibitem{Tsunetoe20}
Y.~{Tsunetoe}, S.~{Mineshige}, K.~{Ohsuga}, T.~{Kawashima}, and K.~{Akiyama},
  ``{Polarization imaging of M 87 jets by general relativistic radiative
  transfer calculation based on GRMHD simulations},''
  \href{https://dx.doi.org/10.1093/pasj/psaa008}{{\em \pasj} {\bfseries 72}
  no.~2, (Apr., 2020) 32}, \href{https://arxiv.org/abs/2002.00954}{{\ttfamily
  arXiv:2002.00954 [astro-ph.HE]}}.

\bibitem{Davelaar23}
J.~{Davelaar}, B.~{Ripperda}, L.~{Sironi}, A.~A. {Philippov}, H.~{Olivares},
  O.~{Porth}, B.~v.~d. {Berg}, T.~{Bronzwaer}, K.~{Chatterjee}, and M.~{Liska},
  ``{Synchrotron Polarization Signatures of Surface Waves in Supermassive Black
  Hole Jets},'' \href{https://dx.doi.org/10.3847/2041-8213/ad0b79}{{\em \apjl}
  {\bfseries 959} no.~1, (Dec., 2023) L3},
  \href{https://arxiv.org/abs/2309.07963}{{\ttfamily arXiv:2309.07963
  [astro-ph.HE]}}.

\bibitem{Blandford77}
R.~D. {Blandford} and R.~L. {Znajek}, ``{Electromagnetic extraction of energy
  from Kerr black holes.},''
  \href{https://dx.doi.org/10.1093/mnras/179.3.433}{{\em \mnras} {\bfseries
  179} (May, 1977) 433--456}.

\bibitem{Blandford82}
R.~D. {Blandford} and D.~G. {Payne}, ``{Hydromagnetic flows from accretion
  disks and the production of radio jets.},''
  \href{https://dx.doi.org/10.1093/mnras/199.4.883}{{\em \mnras} {\bfseries
  199} (June, 1982) 883--903}.

\bibitem{Rybicki79}
G.~B. {Rybicki} and A.~P. {Lightman}, {\em {Radiative processes in
  astrophysics}}.
\newblock 1979.

\bibitem{Zavala03}
R.~T. {Zavala} and G.~B. {Taylor}, ``{A View through Faraday's Fog:
  Parsec-Scale Rotation Measures in Active Galactic Nuclei},''
  \href{https://dx.doi.org/10.1086/374619}{{\em \apj} {\bfseries 589} no.~1,
  (May, 2003) 126--146},
  \href{https://arxiv.org/abs/astro-ph/0302367}{{\ttfamily
  arXiv:astro-ph/0302367 [astro-ph]}}.

\bibitem{Park19}
J.~{Park}, K.~{Hada}, M.~{Kino}, M.~{Nakamura}, H.~{Ro}, and S.~{Trippe},
  ``{Faraday Rotation in the Jet of M87 inside the Bondi Radius: Indication of
  Winds from Hot Accretion Flows Confining the Relativistic Jet},''
  \href{https://dx.doi.org/10.3847/1538-4357/aaf9a9}{{\em \apj} {\bfseries 871}
  no.~2, (Feb., 2019) 257}, \href{https://arxiv.org/abs/1812.08386}{{\ttfamily
  arXiv:1812.08386 [astro-ph.HE]}}.

\bibitem{Hovatta2019}
T.~{Hovatta}, S.~{O'Sullivan}, I.~{Mart{\'\i}-Vidal}, T.~{Savolainen}, and
  A.~{Tchekhovskoy}, ``{Magnetic field at a jet base: extreme Faraday rotation
  in 3C 273 revealed by ALMA},''
  \href{https://dx.doi.org/10.1051/0004-6361/201832587}{{\em \aap} {\bfseries
  623} (Mar., 2019) A111}, \href{https://arxiv.org/abs/1803.09982}{{\ttfamily
  arXiv:1803.09982 [astro-ph.GA]}}.

\bibitem{Broderick09}
A.~E. {Broderick} and A.~{Loeb}, ``{Imaging the Black Hole Silhouette of M87:
  Implications for Jet Formation and Black Hole Spin},''
  \href{https://dx.doi.org/10.1088/0004-637X/697/2/1164}{{\em \apj} {\bfseries
  697} no.~2, (June, 2009) 1164--1179},
  \href{https://arxiv.org/abs/0812.0366}{{\ttfamily arXiv:0812.0366
  [astro-ph]}}.

\bibitem{Belloni10}
T.~M. {Belloni},
  \href{https://dx.doi.org/10.1007/978-3-540-76937-8_3}{``{States and
  Transitions in Black Hole Binaries},''} in {\em Lecture Notes in Physics,
  Berlin Springer Verlag}, T.~{Belloni}, ed., vol.~794, p.~53.
\newblock 2010.

\bibitem{McClintock06}
J.~E. {McClintock} and R.~A. {Remillard},
  \href{https://dx.doi.org/10.48550/arXiv.astro-ph/0306213}{``{Black hole
  binaries},''} in {\em Compact stellar X-ray sources}, W.~H.~G. {Lewin} and
  M.~{van der Klis}, eds., vol.~39, pp.~157--213.
\newblock 2006.

\bibitem{Fender04}
R.~P. {Fender}, T.~M. {Belloni}, and E.~{Gallo}, ``{Towards a unified model for
  black hole X-ray binary jets},''
  \href{https://dx.doi.org/10.1111/j.1365-2966.2004.08384.x}{{\em \mnras}
  {\bfseries 355} no.~4, (Dec., 2004) 1105--1118},
  \href{https://arxiv.org/abs/astro-ph/0409360}{{\ttfamily
  arXiv:astro-ph/0409360 [astro-ph]}}.

\bibitem{Ingram19}
A.~R. {Ingram} and S.~E. {Motta}, ``{A review of quasi-periodic oscillations
  from black hole X-ray binaries: Observation and theory},''
  \href{https://dx.doi.org/10.1016/j.newar.2020.101524}{{\em \nar} {\bfseries
  85} (Sept., 2019) 101524}, \href{https://arxiv.org/abs/2001.08758}{{\ttfamily
  arXiv:2001.08758 [astro-ph.HE]}}.

\bibitem{Majumder26}
S.~{Majumder}, A.~{Kushwaha}, S.~{Singh}, K.~M. {Jayasurya}, S.~{Das}, and
  A.~{Nandi}, ``{Probing the accretion geometry of black hole X-ray binaries: a
  multimission spectro-polarimetric and timing study},''
  \href{https://dx.doi.org/10.1093/mnras/staf1933}{{\em \mnras} {\bfseries 545}
  no.~2, (Jan., 2026) staf1933},
  \href{https://arxiv.org/abs/2506.03774}{{\ttfamily arXiv:2506.03774
  [astro-ph.HE]}}.

\bibitem{Martocchia96}
A.~{Martocchia} and G.~{Matt}, ``{Iron Kalpha line intensity from accretion
  discs around rotating black holes},''
  \href{https://dx.doi.org/10.1093/mnras/282.4.L53}{{\em \mnras} {\bfseries
  282} no.~4, (Oct., 1996) L53--L57}.

\bibitem{Fabian09}
A.~C. {Fabian}, A.~{Zoghbi}, {\em et~al.}, ``{Broad line emission from iron K-
  and L-shell transitions in the active galaxy 1H0707-495},''
  \href{https://dx.doi.org/10.1038/nature08007}{{\em \nat} {\bfseries 459}
  no.~7246, (May, 2009) 540--542}.

\bibitem{Galeev79}
A.~A. {Galeev}, R.~{Rosner}, and G.~S. {Vaiana}, ``{Structured coronae of
  accretion disks.},'' \href{https://dx.doi.org/10.1086/156957}{{\em \apj}
  {\bfseries 229} (Apr., 1979) 318--326}.

\bibitem{Haardt93a}
F.~{Haardt} and L.~{Maraschi}, ``{X-Ray Spectra from Two-Phase Accretion
  Disks},'' \href{https://dx.doi.org/10.1086/173020}{{\em \apj} {\bfseries 413}
  (Aug., 1993) 507}.

\bibitem{Beloborodov99}
A.~M. {Beloborodov}, ``{Plasma Ejection from Magnetic Flares and the X-Ray
  Spectrum of Cygnus X-1},'' \href{https://dx.doi.org/10.1086/311810}{{\em
  \apjl} {\bfseries 510} no.~2, (Jan., 1999) L123--L126},
  \href{https://arxiv.org/abs/astro-ph/9809383}{{\ttfamily
  arXiv:astro-ph/9809383 [astro-ph]}}.

\bibitem{Eardley75}
D.~M. {Eardley}, A.~P. {Lightman}, and S.~L. {Shapiro}, ``{Cygnus X-1: a
  two-temperature accretion disk model which explains the observed hard X-ray
  spectrum.},'' \href{https://dx.doi.org/10.1086/181871}{{\em \apjl} {\bfseries
  199} (Aug., 1975) L153--L155}.

\bibitem{Esin97}
A.~A. {Esin}, J.~E. {McClintock}, and R.~{Narayan}, ``{Advection-Dominated
  Accretion and the Spectral States of Black Hole X-Ray Binaries: Application
  to Nova Muscae 1991},'' \href{https://dx.doi.org/10.1086/304829}{{\em \apj}
  {\bfseries 489} no.~2, (Nov., 1997) 865--889},
  \href{https://arxiv.org/abs/astro-ph/9705237}{{\ttfamily
  arXiv:astro-ph/9705237 [astro-ph]}}.

\bibitem{Haardt93b}
F.~{Haardt} and G.~{Matt}, ``{X-ray polarization in the two-phase model for AGN
  and X-ray binaries},'' \href{https://dx.doi.org/10.1093/mnras/261.2.346}{{\em
  \mnras} {\bfseries 261} no.~2, (Mar., 1993) 346--352}.

\bibitem{Remillard06}
R.~A. {Remillard} and J.~E. {McClintock}, ``{X-Ray Properties of Black-Hole
  Binaries},''
  \href{https://dx.doi.org/10.1146/annurev.astro.44.051905.092532}{{\em \araa}
  {\bfseries 44} no.~1, (Sept., 2006) 49--92},
  \href{https://arxiv.org/abs/astro-ph/0606352}{{\ttfamily
  arXiv:astro-ph/0606352 [astro-ph]}}.

\bibitem{Fragile23b}
P.~C. {Fragile}, K.~{Chatterjee}, A.~{Ingram}, and M.~{Middleton}, ``{The
  luminous, hard state can't be MAD},''
  \href{https://dx.doi.org/10.1093/mnrasl/slad099}{{\em \mnras} {\bfseries 525}
  no.~1, (Oct., 2023) L82--L86},
  \href{https://arxiv.org/abs/2307.08820}{{\ttfamily arXiv:2307.08820
  [astro-ph.HE]}}.

\bibitem{Ingram09}
A.~{Ingram}, C.~{Done}, and P.~C. {Fragile}, ``{Low-frequency quasi-periodic
  oscillations spectra and Lense-Thirring precession},''
  \href{https://dx.doi.org/10.1111/j.1745-3933.2009.00693.x}{{\em \mnras}
  {\bfseries 397} no.~1, (July, 2009) L101--L105},
  \href{https://arxiv.org/abs/0901.1238}{{\ttfamily arXiv:0901.1238
  [astro-ph.SR]}}.

\bibitem{Fragile25b}
P.~C. {Fragile}, D.~A. {Bollimpalli}, J.~D. {Schnittman}, and C.~{Harvey},
  ``{Polarization Signatures of Quasiperiodic Oscillations in Simulated Tilted,
  Truncated Disks},'' \href{https://dx.doi.org/10.3847/1538-4357/adfde1}{{\em
  \apj} {\bfseries 991} no.~1, (Sept., 2025) 80},
  \href{https://arxiv.org/abs/2505.11446}{{\ttfamily arXiv:2505.11446
  [astro-ph.HE]}}.

\bibitem{EHTC_M87_p8}
{Event Horizon Telescope Collaboration}, ``{First M87 Event Horizon Telescope
  Results. VIII. Magnetic Field Structure near The Event Horizon},''
  \href{https://dx.doi.org/10.3847/2041-8213/abe4de}{{\em \apjl} {\bfseries
  910} no.~1, (Mar., 2021) L13},
  \href{https://arxiv.org/abs/2105.01173}{{\ttfamily arXiv:2105.01173
  [astro-ph.HE]}}.

\bibitem{EHTC_M87_p9}
{Event Horizon Telescope Collaboration}, ``{First M87 Event Horizon Telescope
  Results. IX. Detection of Near-horizon Circular Polarization},''
  \href{https://dx.doi.org/10.3847/2041-8213/acff70}{{\em \apjl} {\bfseries
  957} no.~2, (Nov., 2023) L20},
  \href{https://arxiv.org/abs/2311.10976}{{\ttfamily arXiv:2311.10976
  [astro-ph.HE]}}.

\bibitem{EHTC_SgrA_p8}
{Event Horizon Telescope Collaboration}, ``{First Sagittarius A* Event Horizon
  Telescope Results. VIII. Physical Interpretation of the Polarized Ring},''
  \href{https://dx.doi.org/10.3847/2041-8213/ad2df1}{{\em \apjl} {\bfseries
  964} no.~2, (Apr., 2024) L26}.

\bibitem{Moscibrodzka2016}
M.~{Mo{\'s}cibrodzka}, H.~{Falcke}, and H.~{Shiokawa}, ``{General relativistic
  magnetohydrodynamical simulations of the jet in M 87},''
  \href{https://dx.doi.org/10.1051/0004-6361/201526630}{{\em \aap} {\bfseries
  586} (Feb., 2016) A38}, \href{https://arxiv.org/abs/1510.07243}{{\ttfamily
  arXiv:1510.07243 [astro-ph.HE]}}.

\bibitem{Joshi2024}
A.~V. {Joshi}, B.~S. {Prather}, C.-k. {Chan}, M.~{Wielgus}, and C.~F. {Gammie},
  ``{Circular Polarization of Simulated Images of Black Holes},''
  \href{https://dx.doi.org/10.3847/1538-4357/ad5b51}{{\em \apj} {\bfseries 972}
  no.~2, (Sept., 2024) 135}, \href{https://arxiv.org/abs/2406.15653}{{\ttfamily
  arXiv:2406.15653 [astro-ph.HE]}}.

\bibitem{Prather2023}
B.~S. {Prather}, J.~{Dexter}, {\em et~al.}, ``{Comparison of Polarized
  Radiative Transfer Codes Used by the EHT Collaboration},''
  \href{https://dx.doi.org/10.3847/1538-4357/acc586}{{\em \apj} {\bfseries 950}
  no.~1, (June, 2023) 35}, \href{https://arxiv.org/abs/2303.12004}{{\ttfamily
  arXiv:2303.12004 [astro-ph.HE]}}.

\bibitem{Davelaar2019}
J.~{Davelaar}, H.~{Olivares}, O.~{Porth}, T.~{Bronzwaer}, M.~{Janssen},
  F.~{Roelofs}, Y.~{Mizuno}, C.~M. {Fromm}, H.~{Falcke}, and L.~{Rezzolla},
  ``{Modeling non-thermal emission from the jet-launching region of M 87 with
  adaptive mesh refinement},''
  \href{https://dx.doi.org/10.1051/0004-6361/201936150}{{\em \aap} {\bfseries
  632} (Dec., 2019) A2}, \href{https://arxiv.org/abs/1906.10065}{{\ttfamily
  arXiv:1906.10065 [astro-ph.HE]}}.

\bibitem{Chael2019}
A.~{Chael}, R.~{Narayan}, and M.~D. {Johnson}, ``{Two-temperature, Magnetically
  Arrested Disc simulations of the jet from the supermassive black hole in
  M87},'' \href{https://dx.doi.org/10.1093/mnras/stz988}{{\em \mnras}
  {\bfseries 486} no.~2, (June, 2019) 2873--2895},
  \href{https://arxiv.org/abs/1810.01983}{{\ttfamily arXiv:1810.01983
  [astro-ph.HE]}}.

\bibitem{EHTC_SgrA_p6}
{Event Horizon Telescope Collaboration}, ``{First Sagittarius A* Event Horizon
  Telescope Results. VI. Testing the Black Hole Metric},''
  \href{https://dx.doi.org/10.3847/2041-8213/ac6756}{{\em \apjl} {\bfseries
  930} no.~2, (May, 2022) L17}.

\bibitem{Gravity2018}
{GRAVITY Collaboration}, R.~{Abuter}, {\em et~al.}, ``{Detection of orbital
  motions near the last stable circular orbit of the massive black hole
  SgrA*},'' \href{https://dx.doi.org/10.1051/0004-6361/201834294}{{\em \aap}
  {\bfseries 618} (Oct., 2018) L10},
  \href{https://arxiv.org/abs/1810.12641}{{\ttfamily arXiv:1810.12641
  [astro-ph.GA]}}.

\bibitem{Wielgus2022polar}
M.~{Wielgus}, M.~{Moscibrodzka}, J.~{Vos}, Z.~{Gelles}, I.~{Mart{\'\i}-Vidal},
  J.~{Farah}, N.~{Marchili}, C.~{Goddi}, and H.~{Messias}, ``{Orbital motion
  near Sagittarius A$^{*}$ . Constraints from polarimetric ALMA
  observations},'' \href{https://dx.doi.org/10.1051/0004-6361/202244493}{{\em
  \aap} {\bfseries 665} (Sept., 2022) L6},
  \href{https://arxiv.org/abs/2209.09926}{{\ttfamily arXiv:2209.09926
  [astro-ph.HE]}}.

\bibitem{Yuan2014}
F.~{Yuan} and R.~{Narayan}, ``{Hot Accretion Flows Around Black Holes},''
  \href{https://dx.doi.org/10.1146/annurev-astro-082812-141003}{{\em \araa}
  {\bfseries 52} (Aug., 2014) 529--588},
  \href{https://arxiv.org/abs/1401.0586}{{\ttfamily arXiv:1401.0586
  [astro-ph.HE]}}.

\bibitem{EHT_M87_2025}
{Event Horizon Telescope Collaboration}, K.~{Akiyama}, {\em et~al.},
  ``{Horizon-scale variability of M87* from 2017-2021 EHT observations},''
  \href{https://dx.doi.org/10.1051/0004-6361/202555855}{{\em \aap} {\bfseries
  704} (Dec., 2025) A91}, \href{https://arxiv.org/abs/2509.24593}{{\ttfamily
  arXiv:2509.24593 [astro-ph.HE]}}.

\bibitem{Wielgus2022EHT}
M.~{Wielgus}, N.~{Marchili}, {\em et~al.}, ``{Millimeter Light Curves of
  Sagittarius A* Observed during the 2017 Event Horizon Telescope Campaign},''
  \href{https://dx.doi.org/10.3847/2041-8213/ac6428}{{\em \apjl} {\bfseries
  930} no.~2, (Jan., 2022) L19},
  \href{https://arxiv.org/abs/2207.06829}{{\ttfamily arXiv:2207.06829
  [astro-ph.HE]}}.

\bibitem{Wielgus2024}
M.~{Wielgus}, S.~{Issaoun}, I.~{Mart{\'\i}-Vidal}, R.~{Emami},
  M.~{Moscibrodzka}, C.~D. {Brinkerink}, C.~{Goddi}, and E.~{Fomalont}, ``{The
  internal Faraday screen of Sagittarius A*},''
  \href{https://dx.doi.org/10.1051/0004-6361/202347772}{{\em \aap} {\bfseries
  682} (Feb., 2024) A97}, \href{https://arxiv.org/abs/2308.11712}{{\ttfamily
  arXiv:2308.11712 [astro-ph.HE]}}.

\bibitem{Dhruv2025}
V.~{Dhruv}, B.~{Prather}, M.~{Chandra}, A.~V. {Joshi}, and C.~F. {Gammie},
  ``{Electromagnetic Observables of Weakly Collisional Black Hole Accretion},''
  \href{https://dx.doi.org/10.3847/2041-8213/ae1236}{{\em \apjl} {\bfseries
  993} no.~1, (Nov., 2025) L33},
  \href{https://arxiv.org/abs/2510.11365}{{\ttfamily arXiv:2510.11365
  [astro-ph.HE]}}.

\bibitem{Issaoun2022}
S.~{Issaoun}, M.~{Wielgus}, {\em et~al.}, ``{Resolving the Inner Parsec of the
  Blazar J1924-2914 with the Event Horizon Telescope},''
  \href{https://dx.doi.org/10.3847/1538-4357/ac7a40}{{\em \apj} {\bfseries 934}
  no.~2, (Aug., 2022) 145}, \href{https://arxiv.org/abs/2208.01662}{{\ttfamily
  arXiv:2208.01662 [astro-ph.HE]}}.

\bibitem{Jorstad2023}
S.~{Jorstad}, M.~{Wielgus}, {\em et~al.}, ``{The Event Horizon Telescope Image
  of the Quasar NRAO 530},''
  \href{https://dx.doi.org/10.3847/1538-4357/acaea8}{{\em \apj} {\bfseries 943}
  no.~2, (Feb., 2023) 170}, \href{https://arxiv.org/abs/2302.04622}{{\ttfamily
  arXiv:2302.04622 [astro-ph.HE]}}.

\bibitem{Paraschos2024}
G.~F. {Paraschos}, J.-Y. {Kim}, {\em et~al.}, ``{Ordered magnetic fields around
  the 3C 84 central black hole},''
  \href{https://dx.doi.org/10.1051/0004-6361/202348308}{{\em \aap} {\bfseries
  682} (Feb., 2024) L3}, \href{https://arxiv.org/abs/2402.00927}{{\ttfamily
  arXiv:2402.00927 [astro-ph.HE]}}.

\bibitem{Roeder2025}
J.~{R{\"o}der}, M.~{Wielgus}, {\em et~al.}, ``{A multifrequency study of
  sub-parsec jets with the Event Horizon Telescope},''
  \href{https://dx.doi.org/10.1051/0004-6361/202452600}{{\em \aap} {\bfseries
  695} (Mar., 2025) A233}, \href{https://arxiv.org/abs/2501.05518}{{\ttfamily
  arXiv:2501.05518 [astro-ph.HE]}}.

\bibitem{Chen2024}
C.-T.~J. {Chen}, I.~{Liodakis}, {\em et~al.}, ``{X-Ray and Multiwavelength
  Polarization of Mrk 501 from 2022 to 2023},''
  \href{https://dx.doi.org/10.3847/1538-4357/ad63a1}{{\em \apj} {\bfseries 974}
  no.~1, (Oct., 2024) 50}, \href{https://arxiv.org/abs/2407.11128}{{\ttfamily
  arXiv:2407.11128 [astro-ph.HE]}}.

\bibitem{Barnier24}
S.~{Barnier} and C.~{Done}, ``{Making the Invisible Visible: Magnetic Fields in
  Accretion Flows Revealed by X-Ray Polarization},''
  \href{https://dx.doi.org/10.3847/1538-4357/ad9277}{{\em \apj} {\bfseries 977}
  no.~2, (Dec., 2024) 201}, \href{https://arxiv.org/abs/2404.12815}{{\ttfamily
  arXiv:2404.12815 [astro-ph.HE]}}.

\bibitem{Zhao24}
Q.-C. {Zhao}, L.~{Tao}, {\em et~al.}, ``{The First Polarimetric View on
  Quasiperiodic Oscillations in a Black Hole X-Ray Binary},''
  \href{https://dx.doi.org/10.3847/2041-8213/ad1e6c}{{\em \apjl} {\bfseries
  961} no.~2, (Feb., 2024) L42},
  \href{https://arxiv.org/abs/2401.08970}{{\ttfamily arXiv:2401.08970
  [astro-ph.HE]}}.

\end{thebibliography}
\providecommand{\href}[2]{#2}\begingroup\raggedright\endgroup


\end{document}